\documentclass[%
reprint,
superscriptaddress,
 amsmath,amssymb,
 aps,
]{revtex4-2}

\usepackage{graphicx}
\usepackage{stackengine}
\usepackage{dcolumn}
\usepackage{bm}

\usepackage[caption=false]{subfig}
\usepackage{siunitx, soul, xcolor}
\usepackage{adjustbox}
\usepackage{gensymb}

\usepackage{xparse,xcoffins}

\ExplSyntaxOn
\NewCoffin\imagecoffin
\NewCoffin\labelcoffin

\keys_define:nn { miguel/label }
 {
  label   .tl_set:N = \l_miguel_label_tl,
  labelbox .bool_set:N = \l_miguel_label_box_bool,
  labelbox .default:n = true,
  fontsize .tl_set:N = \l_miguel_label_size_tl,
  fontsize .initial:n = \footnotesize,
  pos .choice:,
  pos/nw .code:n = \tl_set:Nn \l_miguel_label_pos_tl { left,up },
  pos/ne .code:n = \tl_set:Nn \l_miguel_label_pos_tl { right,up },
  pos/sw .code:n = \tl_set:Nn \l_miguel_label_pos_tl { left,down },
  pos/se .code:n = \tl_set:Nn \l_miguel_label_pos_tl { right,down },
  pos/n .code:n = \tl_set:Nn \l_miguel_label_pos_tl { hc,up },
  pos/w .code:n = \tl_set:Nn \l_miguel_label_pos_tl { left,vc },
  pos/s .code:n = \tl_set:Nn \l_miguel_label_pos_tl { hc,down },
  pos/e .code:n = \tl_set:Nn \l_miguel_label_pos_tl { right,vc },
  pos .initial:n = nw,
  unknown .code:n   = \clist_put_right:Nx \l_miguel_label_clist
                       { \l_keys_key_tl = \exp_not:n { #1 } }
 }
\clist_new:N \l_miguel_label_clist
\box_new:N \l_miguel_label_box
\box_new:N \l_miguel_label_image_box

\NewDocumentCommand{\xincludegraphics}{O{}m}
 {
  \group_begin:
  \tl_clear:N \l_miguel_label_tl
  \clist_clear:N \l_miguel_label_clist
  \keys_set:nn { miguel/label } { #1 }
  \tl_if_empty:NTF \l_miguel_label_tl
   {
    \miguel_includegraphics:Vn \l_miguel_label_clist { #2 }
   }
   {
    \SetHorizontalCoffin\imagecoffin
     {
      \miguel_includegraphics:Vn \l_miguel_label_clist { #2 }
     }
    \SetHorizontalCoffin\labelcoffin
     {
      \raisebox{\depth}
       {
        \bool_if:NTF \l_miguel_label_box_bool
         { \fcolorbox{white}{white}{\l_miguel_label_size_tl\l_miguel_label_tl} }
         { \l_miguel_label_size_tl\l_miguel_label_tl }
       }
     }
    \SetVerticalPole\imagecoffin{left}{-0pt+\CoffinWidth\labelcoffin/2}
    \SetVerticalPole\imagecoffin{right}{\Width-3pt-\CoffinWidth\labelcoffin/2}
    \SetHorizontalPole\imagecoffin{up}{\Height+8pt-\CoffinHeight\labelcoffin/2}
    \SetHorizontalPole\imagecoffin{down}{3pt+\CoffinHeight\labelcoffin/2}
    \use:x{\JoinCoffins\imagecoffin[\l_miguel_label_pos_tl]\labelcoffin[vc,hc]} 
    \TypesetCoffin\imagecoffin
   }
   \group_end:
 }
\NewDocumentCommand{\setlabel}{m}
 {
  \keys_set:nn { miguel/label } { #1 }
 }

\cs_new_protected:Nn \miguel_includegraphics:nn
 {
  \includegraphics[#1]{#2}
 }
\cs_generate_variant:Nn \miguel_includegraphics:nn { V }

\ExplSyntaxOff

\begin{document}
\preprint{APS/123-QED}

\title{Electrical Side-Gate Control of Anisotropic Magnetoresistance and Magnetic Anisotropy in a Composite Multiferroic}

\author{Katherine Johnson}
\email{robinson.1971@osu.edu}
\affiliation{Department of Physics, The Ohio State University, Columbus, Ohio 43210, United States}

\author{Michael Newburger}
\affiliation{Materials and Manufacturing Directorate, Air Force Research Laboratory, Wright-Patterson Air Force Base, Ohio 45433, USA}
\author{Michael Page}
\affiliation{Materials and Manufacturing Directorate, Air Force Research Laboratory, Wright-Patterson Air Force Base, Ohio 45433, USA}
\author{Roland K. Kawakami}
\affiliation{Department of Physics, The Ohio State University, Columbus, Ohio 43210, United States}

\

\begin{abstract}

Composite multiferroics consisting of a ferroelectric material interfaced with a ferromagnetic material can function above room temperature and exhibit improved magnetoelectric (ME) coupling compared to single-phase multiferroic materials, making them desirable for applications in energy-efficient electronic devices.   
In this study, we demonstrate electrical side-gate control of magnetoresistance and magnetic anisotropy in single-crystalline ferromagnetic Fe$_{0.75}$Co$_{0.25}$ thin films grown on ferroelectric PMN-PT (001) substrates by molecular beam epitaxy. 
Fe$_{0.75}$Co$_{0.25}$ is selected due to its large magnetoelastic coupling and low magnetic damping. We find that the magnetoresistance curves of patterned Fe$_{0.75}$Co$_{0.25}$ films are controlled by voltages applied to electrostatic side gates. 
Angle-dependent magnetoresistance scans reveal that the origin of this effect is strain-mediated variation of the magnetic anisotropy due to piezoelectric effects in the PMN-PT. This electrical control of magnetic properties could serve as a building block for future magnetoelectronic and magnonic devices.
\end{abstract}

\flushbottom
\maketitle
\thispagestyle{empty}

\section{Introduction}

Multiferroics are a thriving field of interest due to their potential for enhanced performance in a wide range of applications including ultra-low power and high-density memory and logic, high-frequency devices, sensors, inductors, actuators, energy harvesters, miniature antennas, and terahertz emitters \cite{liu_voltage_2014,liang_magnetoelectric_2021,pradhan_magnetoelectric_2020,chumak_magnon_2015,wang_applications_2016}.
Magnetic microwave devices provide an example of the possible gains in energy efficiency and miniaturization that could be achieved with multiferroics~\cite{srinivasan_ferrite-piezoelectric_2006,liu_voltage_2014}.
In most commercial devices, control of the magnetization requires the use of electromagnets which are large, slow, and use a significant amount of power.  
Multiferroics could circumvent this issue by utilizing coupled ferroic properties, i.e. ferroelectric and ferromagnetic properties \cite{liu_voltage_2014,liang_magnetoelectric_2021}, to enable faster and more efficient operation through electrical voltage control of magnetic properties~\cite{matsukura_control_2015,spaldin_advances_2019,wang_applications_2016}. 

The multiferroic coupling is typically present in two forms, either in a single material (``type-I'') or in a composite heterostructure (``type-II'').
While type-I multiferroics have garnered substantial scientific interest, they are typically limited for applications because either the ferroelectricity or magnetism is weak due to competing interactions~\cite{hill_why_2000,fiebig_revival_2005,spaldin_renaissance_2005,eerenstein_multiferroic_2006}.
Type-II multiferroics may provide a better path toward applications because ferroelectric materials and ferromagnetic materials with desired characteristics can be selected and heterogeneously integrated to leverage the best of both worlds~\cite{liu_voltage_2014}. The challenge comes in finding compatible materials and understanding the source of the magnetoelectric coupling. 

\begin{figure}[h]
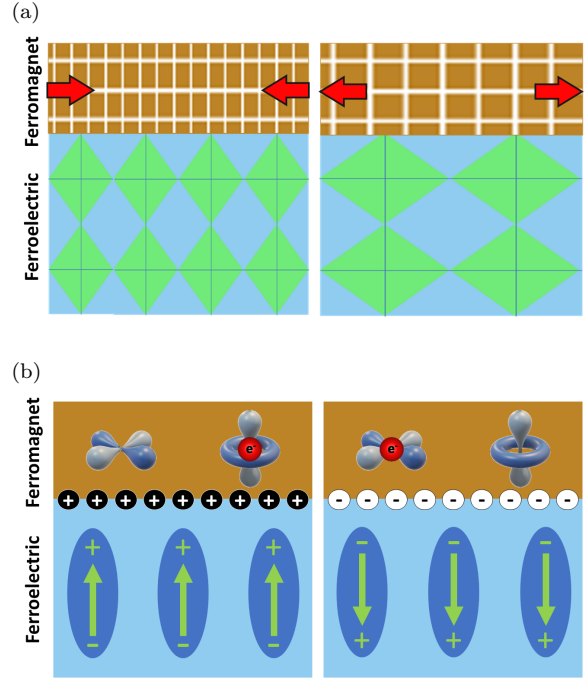

\subfloat{\xincludegraphics[width=0.43\textwidth,label=(a)]{figures/mecoupling_strain.png}\label{fig:ME_strain}}\hfill
\subfloat{\xincludegraphics[width=0.43\textwidth,label=(b)]{figures/mecoupling_charge.png}\label{fig:ME_charge}}\hfill

\vspace{-0.01\textwidth}
\caption{Two different magneto-electric coupling mechanisms:
  (a) strain-mediated,
  (b) interfacial charge-mediated.
  }
  \label{fig:ME_coupling}
\end{figure}

There are multiple magnetoelectric couplings in a composite multiferroic that can occur. The two main ones are strain (Fig.~\ref{fig:ME_strain}) and interfacial charge accumulation (Fig.~\ref{fig:ME_charge}). The strain occurs due to the piezoelectric component where applying an electric field causes a strain in the crystal structure of the ferroelectric. This strain, when coupled to a magnetostrictive material, can change the magnetic properties of the material. The other mechanism is polarization-mediated coupling, which is caused by charge accumulation at the interface of the multiferroic heterostructure. This, in turn, causes a change of the density of states of the magnetic material, which affect the magnetic properties~\cite{maruyama_large_2009}. Fig.~\ref{fig:ME_charge} shows a conceptual drawing describing the effect of the electric field of the charge accumulation on the electron filling of the 3d orbitals in the ferromagnet. This effect has a short length scale, inferred to be a few atomic layers. However, according to Jia \textit{et al.}~\cite{jia_mechanism_2014}, for a ferromagnet, the charge accumulation can cause a surface spiral spin density to form, resulting in longer range spin rearrangement than expected from electrostatic charge screening. 
Some experimental works claim this to be the main magnetoelectric coupling effect in their systems~\cite{zhou_long-range_2018}.
However, separating the types of magnetoelectric coupling occurring in a system is not trivial. 
In fact, sometimes multiple types of coupling can occur in the same system simultaneously~\cite{liang_magnetoelectric_2021,zhou_long-range_2018,hu_multiferroic_2015,wang_multiferroic_2010}. This complicates understanding the system for application in more complex device structures, such as surface acoustic wave (SAW) devices~\cite{weiler_elastically_2011,thevenard_surface-acoustic-wave-driven_2014,gowtham_mechanical_2016,labanowski_power_2016,li_spin_2017,yamamoto_interaction_2022}.  

In this paper, we report the electrical side-gate control of anisotropic magnetoresistance and magnetic anisotropy in a type-II (composite) multiferroic consisting of ferromagnetic Fe$_{0.75}$Co$_{0.25}$ and ferroelectric {Pb}({Mg}$_{1/3}${Nb}$_{2/3}$)$_{0.63}${Ti}$_{0.37}${O}$_3$ (PMN-PT, composition verified using EDX).
 Fe$_{0.75}$Co$_{0.25}$ is a low-damping ferromagnet allowing for longer lifetimes of spin waves~\cite{schwienbacher_magnetoelasticity_2019,lee_metallic_2017}.
Fe$_{0.75}$Co$_{0.25}$ is also magnetoelastic, meaning the magnetic properties are greatly influenced by strain~\cite{zuberek_magnetostriction_2000}. 
Conveniently, Fe$_{0.75}$Co$_{0.25}$ also forms a body centered cubic crystal structure with a lattice constant of ($\sim 0.286$\,nm) which closely matches multiple ferroelectric materials, allowing for epitaxial growth. PMN-PT is a well-known relaxor ferroelectric with excellent piezoelectric response and good acoustoelectric conversion efficiency, making it a good material for study applications like SAW devices.
In this study, we develop the epitaxial growth of (001)-oriented Fe$_{0.75}$Co$_{0.25}$ films on PMN-PT using molecular beam epitaxy (MBE). 
We measure the magnetoresistance of a patterned Fe$_{0.75}$Co$_{0.25}$ device channel and apply electric fields to the underlying PMN-PT using a lateral side-gate geometry.
Unlike vertical back-gate geometries that are typically employed in type-II multiferroic studies~\cite{dong_piezoelectric_2014,zhou_long-range_2018,liang_magnetoelectric_2021,yang_angle-dependent_2021,wang_non-volatile_2023}, the lateral geometry is more convenient for some device structures (e.g.~interdigitated transducers)~\cite{weiler_elastically_2011,thevenard_surface-acoustic-wave-driven_2014,gowtham_mechanical_2016,labanowski_power_2016,li_spin_2017,yamamoto_interaction_2022} and could produce equivalent electric fields with orders of magnitude less voltage.
Measuring the channel resistance as a function of magnetic field shows signatures of anisotropic magnetoresistance (AMR), and the magnetoresistance curves are found to be reversibly tunable by gate voltage.
Angle-dependent magnetoresistance scans (i.e., rotating the magnetic field with a fixed magnitude) confirm the AMR and identify gate control of the uniaxial magnetic anisotropy as responsible for the observed tunability in the AMR curves.
Further, an epitaxial Cr interlayer is inserted (i.e.~Fe$_{0.75}$Co$_{0.25}$/Cr/PMN-PT) to investigate whether the magnetoelectric coupling originates from a strain-mediated effect or a charge-mediated effect, and the results support a predominantly strain-driven mechanism.

\section{Material Synthesis and Characterization}

\begin{figure}[h!]
\subfloat{\xincludegraphics[width=0.23\textwidth,label=(a)]{figures/PMN-PT_draft2.png}\label{fig:rheed-PMNPT}}\hfill
\subfloat{\xincludegraphics[width=0.23\textwidth,label=(b)]{figures/RHEEDFeCo_draft2.png}\label{fig:rheed-FeCo}}\hfill
\subfloat{\xincludegraphics[width=0.23\textwidth,label=(c)]{figures/Cr-PMNPT_draft2.png}\label{fig:rheed-CrPMNPT}}\hfill
\subfloat{\xincludegraphics[width=0.23\textwidth,label=(d)]{figures/RHEEDFCC_draft2.png}\label{fig:rheed-FeCoCr}}\hfill
\subfloat{\xincludegraphics[width=0.48\textwidth,label=(e)]{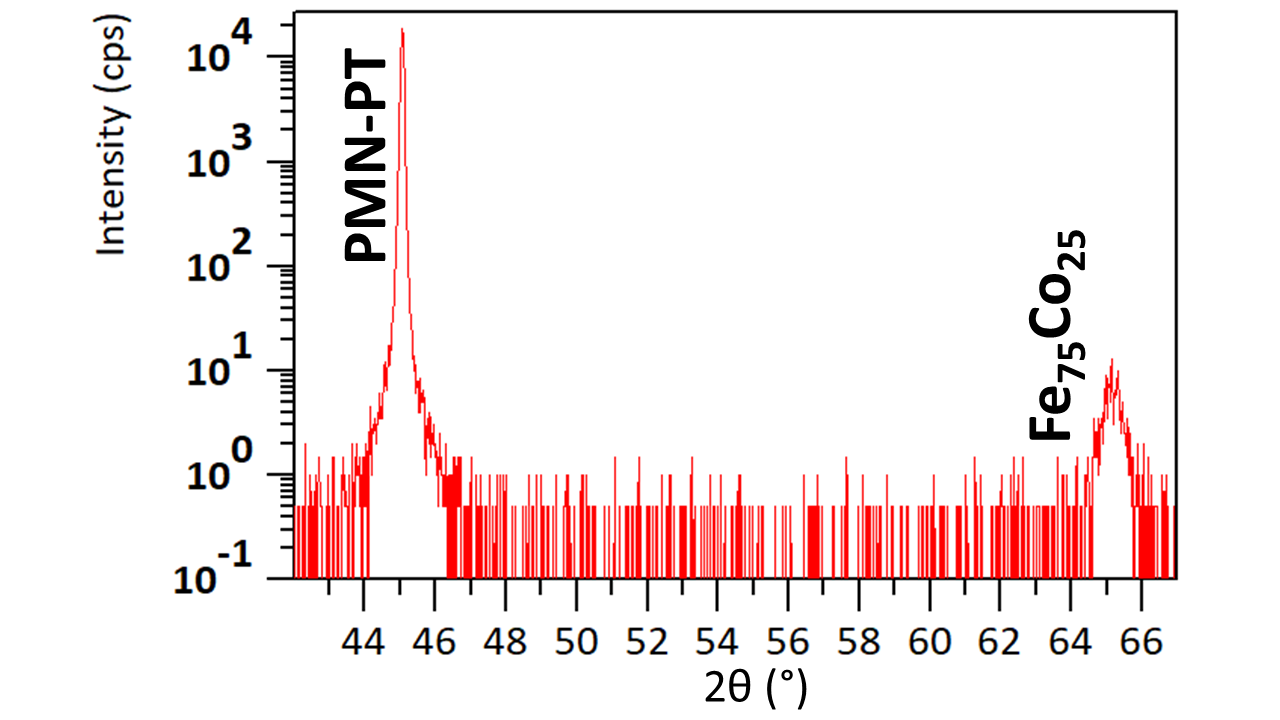}\label{fig:xrdfeco}}\hfill
\subfloat{\xincludegraphics[width=0.48\textwidth,label=(f)]{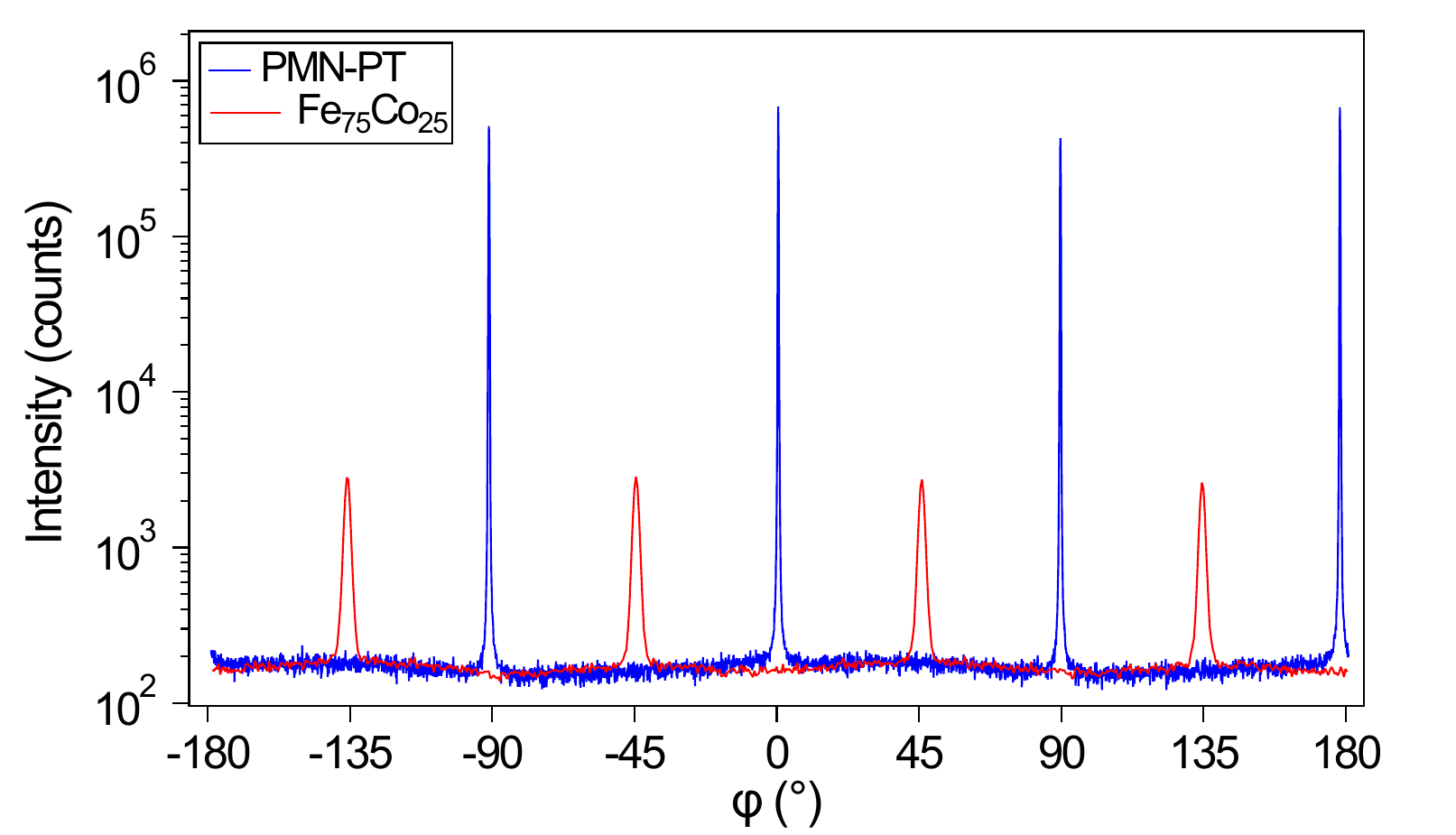}\label{fig:fecophiscan}}\hfill
\caption{RHEED patterns and X-Ray Diffraction of the growths on (a) PMN-PT substrate, (b) 10\,nm Fe$_{0.75}$Co$_{0.25}$ on the PMN-PT substrate, (c) 7\,nm of Cr after anneal on PMN-PT substrate, and (d) 10\,nm Fe$_{0.75}$Co$_{0.25}$ on 7\,nm Cr buffer layer on PMN-PT substrate. XRD (f) shows the PMN-PT peak and the Fe$_{0.75}$Co$_{0.25}$ peak, and the phi scan (e) shows the rotation of Fe$_{0.75}$Co$_{0.25}$ growth on PMN-PT substrate.}
  \label{fig:rheed}
\end{figure}

Samples are grown in an MBE chamber with a base pressure of $8\times 10^{-10}$ Torr. The (001)-oriented PMN-PT substrates are prepared by sonication in acetone and isopropyl alcohol for five minutes each.
The substrates are then introduced into the MBE chamber and annealed at $500\degree$C. For the MgO(001) substrates, they are prepared by sonication in acetone, isopropyl alcohol, and water for five minutes each, and then \textit{in-situ} annealed $500\degree$C. 
Elemental Fe, Co, and Cr are deposited from effusion cells with typical growth rates of $0.10 - 0.35$ nm/min as measured by a quartz deposition monitor. The Fe$_{0.75}$Co$_{0.25}$ films are deposited onto substrates held at a temperature of $250\degree$C and \textit{in situ} reflection high-energy electron diffraction (RHEED) monitors the surface structure during the growth.

Figure~\ref{fig:rheed-PMNPT} displays RHEED of the PMN-PT substrate, which shows the surface of the PMN-PT is flat and a good interface for epitaxial growth. Figure~\ref{fig:rheed-CrPMNPT} shows the RHEED of 7\,nm of Cr on PMN-PT, grown as an interlayer for device C. The Cr/PMN-PT RHEED indicates a crystalline and two-dimensional growth for a good interface for Fe$_{0.75}$Co$_{0.25}$ growth. Figures~\ref{fig:rheed-FeCo} and \ref{fig:rheed-FeCoCr} show the RHEED patterns of 10\,nm films of Fe$_{0.75}$Co$_{0.25}$ grown on PMN-PT substrates and on PMN-PT substrates with a 7\,nm Cr buffer layer, respectively. The streaky patterns indicate two-dimensional growth with finite terrace width. 
The streak spacings in the RHEED patterns help establish the in-plane epitaxial relationships among the bcc Fe$_{0.75}$Co$_{0.25}$ film, bcc Cr film, and PMN-PT substrate. Because the streak spacing of Fe$_{0.75}$Co$_{0.25}$ (Figure~\ref{fig:rheed-FeCo}) is roughly double that of PMN-PT (Figure~\ref{fig:rheed-PMNPT}), this indicates a 45$\degree$ rotation of the unit cell as their lattice constants differ by a factor of $\sim$$\sqrt{2}$ ($a_{\text{FeCo}}$ $\sim$0.286 nm \cite{lee_metallic_2017} and $a_{\text{PMNPT}} = 0.4024$\,nm).
These epitaxial relationships are confirmed by x-ray diffraction. The $\theta-2\theta$ scan (Figure~\ref{fig:xrdfeco}) only shows peaks corresponding to (001)-oriented PMN-PT and (001)-oriented Fe$_{0.75}$Co$_{0.25}$. For the azimuthal in-plane rotation (i.e. phi scan) in Figure~\ref{fig:fecophiscan}, the peaks for the Fe$_{0.75}$Co$_{0.25}$ film (red curve) and PMN-PT substrate (blue curve) collected at different polar angles are offset by 45$\degree$, which confirms the 45$\degree$ in-plane rotation between the film and the substrate.


\begin{figure*}[t!]
\subfloat{\xincludegraphics[width=0.15\textwidth,label=(a)]{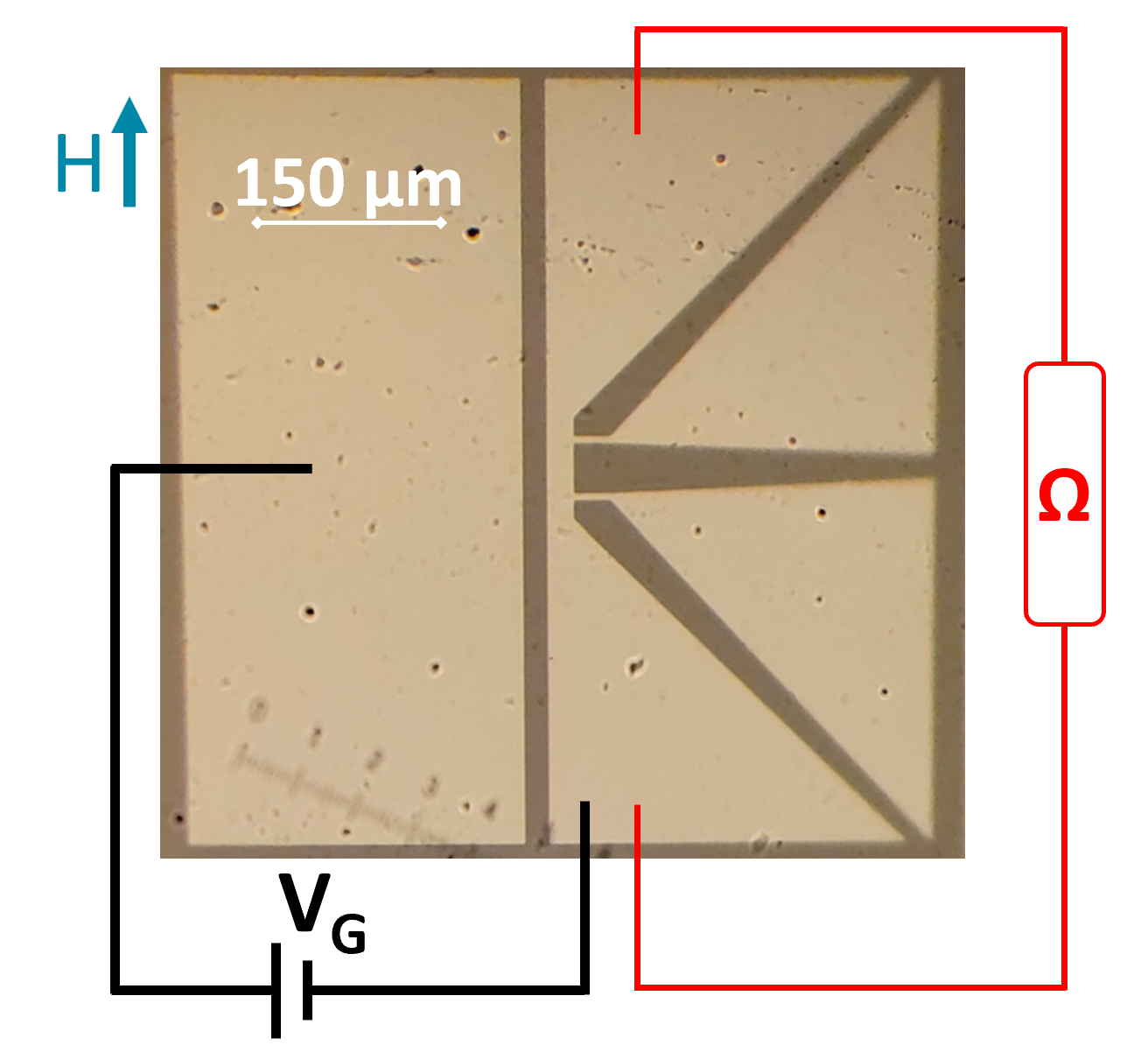}\label{fig:DevDiaA}}\hfill
\subfloat{\xincludegraphics[width=0.33\textwidth,label=(b)]{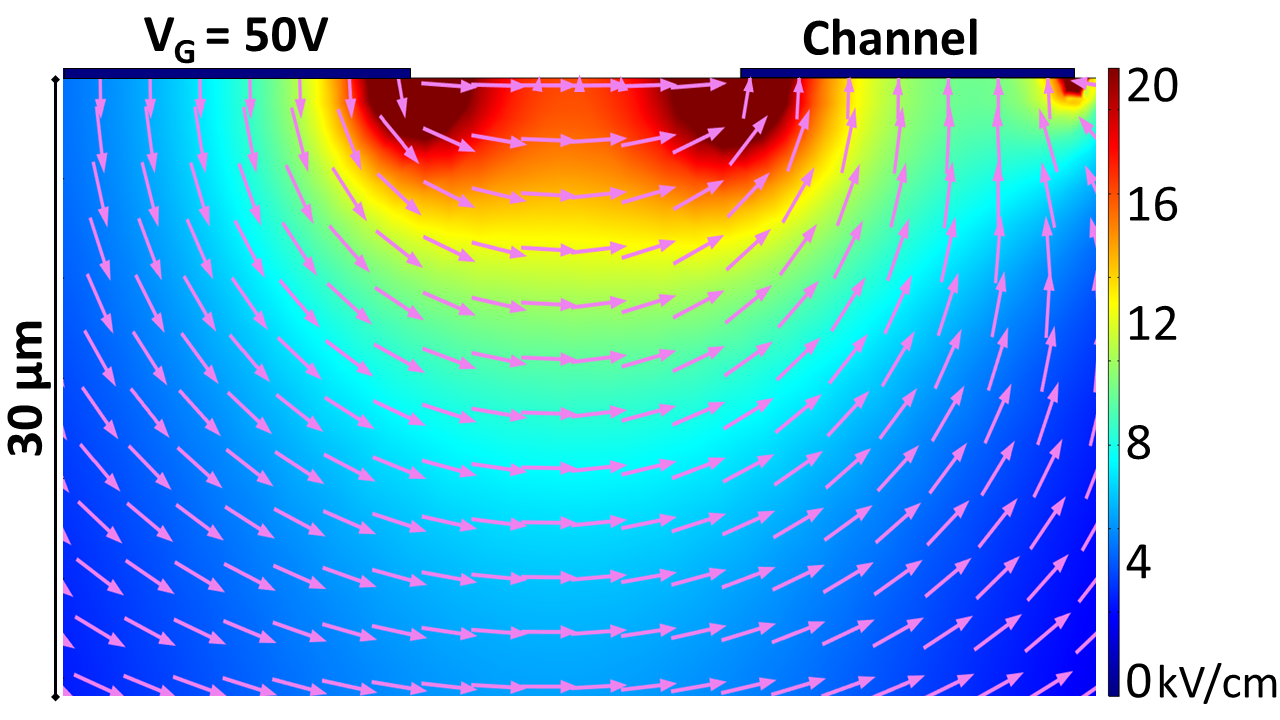}\label{fig:DevcomsolA}}\hfill
\subfloat{\xincludegraphics[width=0.15\textwidth,label=(e)]{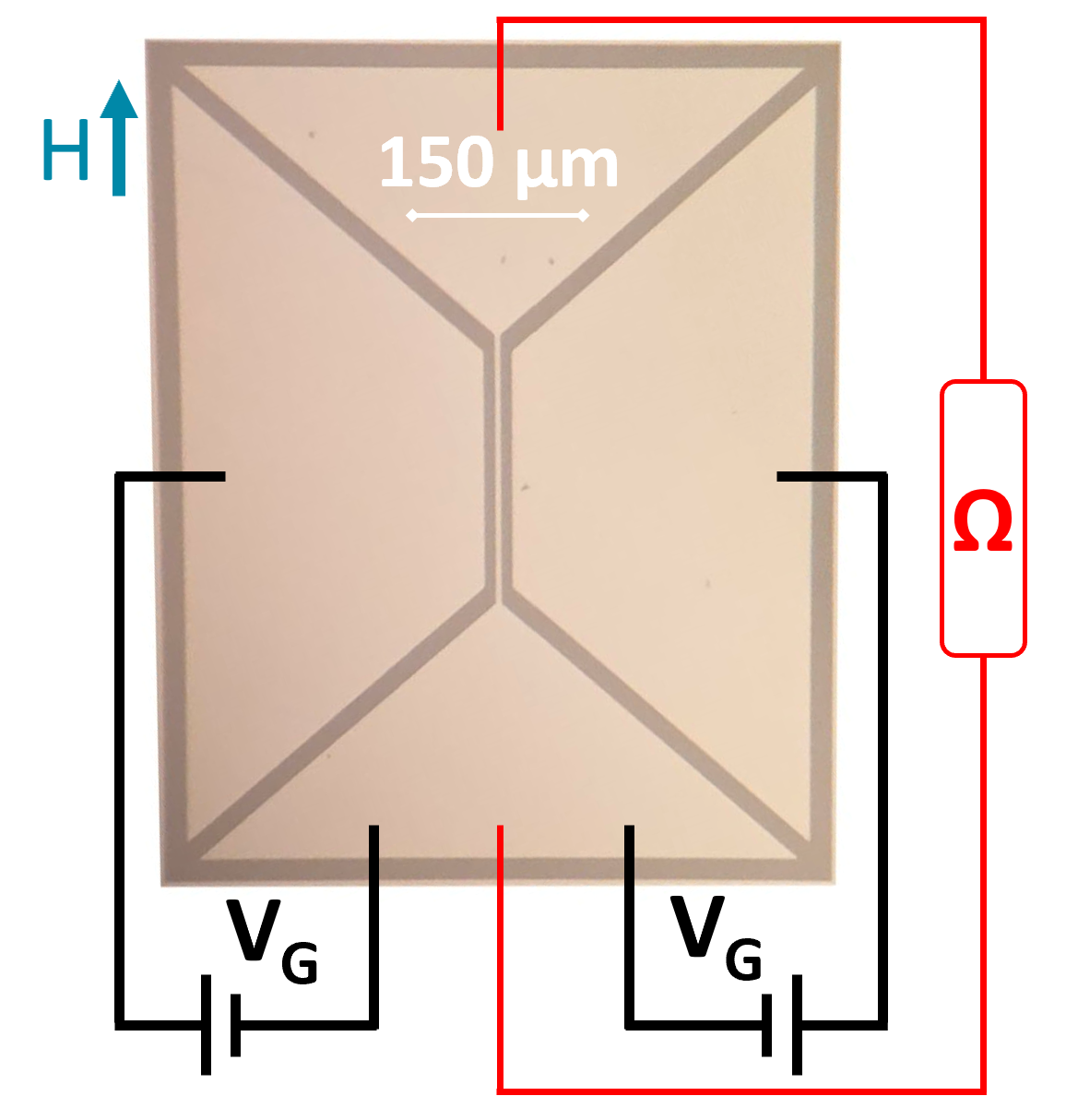}\label{fig:upsweepDevA}}\hfill
\subfloat{\xincludegraphics[width=0.33\textwidth,label=(f)]{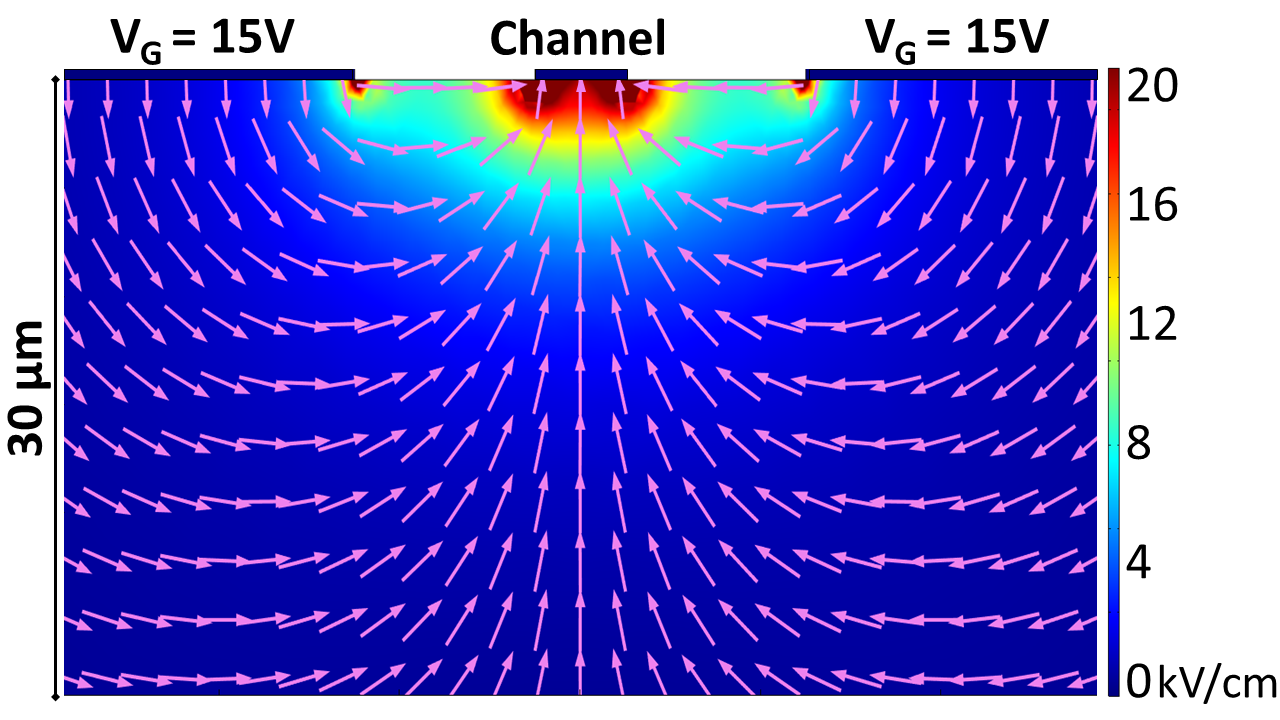}\label{fig:downsweepDevA}}\hfill
\subfloat{\xincludegraphics[width=0.44\textwidth,label=(c)]{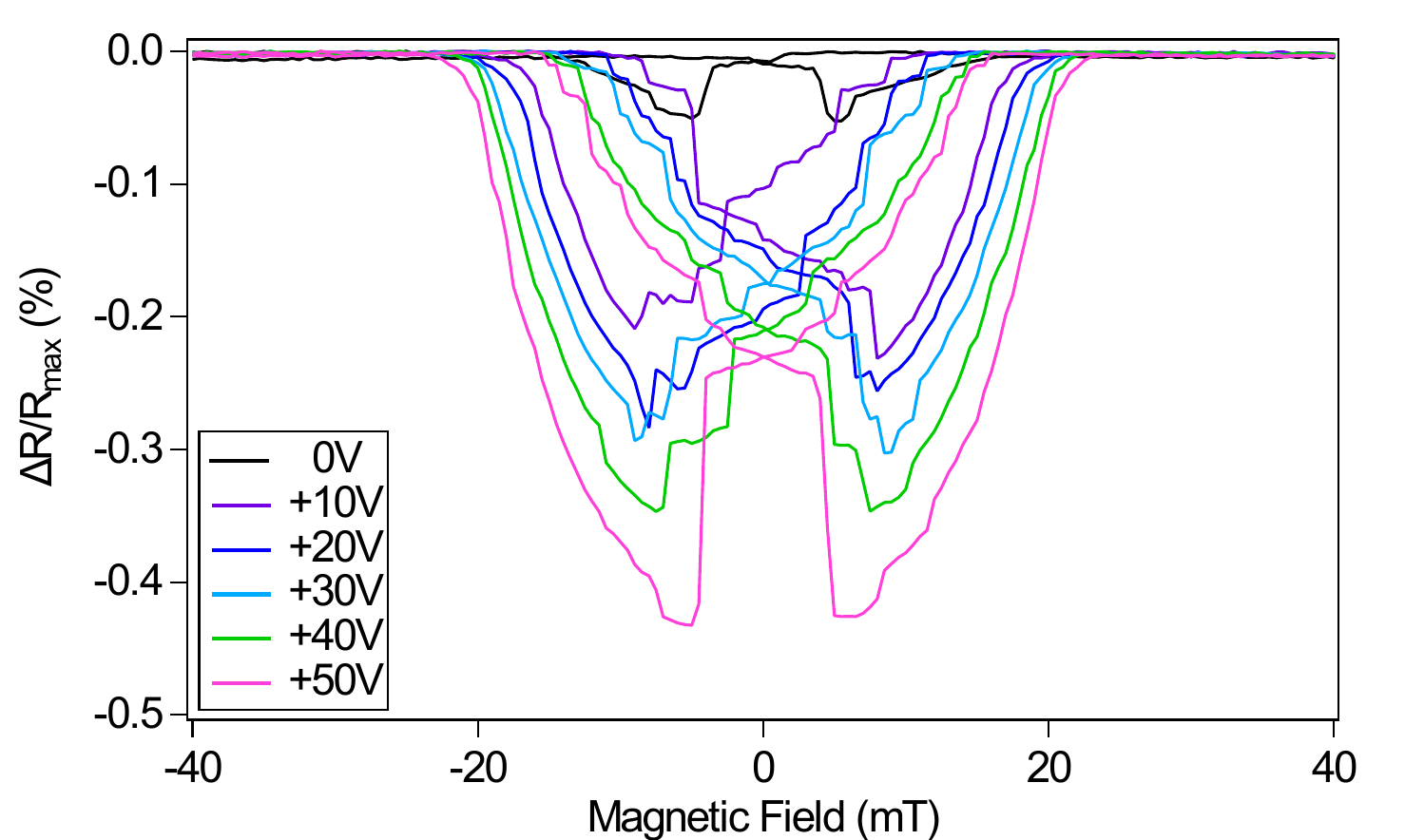}\label{fig:DevDiaB}}\hspace{0.1cm}
\subfloat{\xincludegraphics[width=0.44\textwidth,label=(g)]{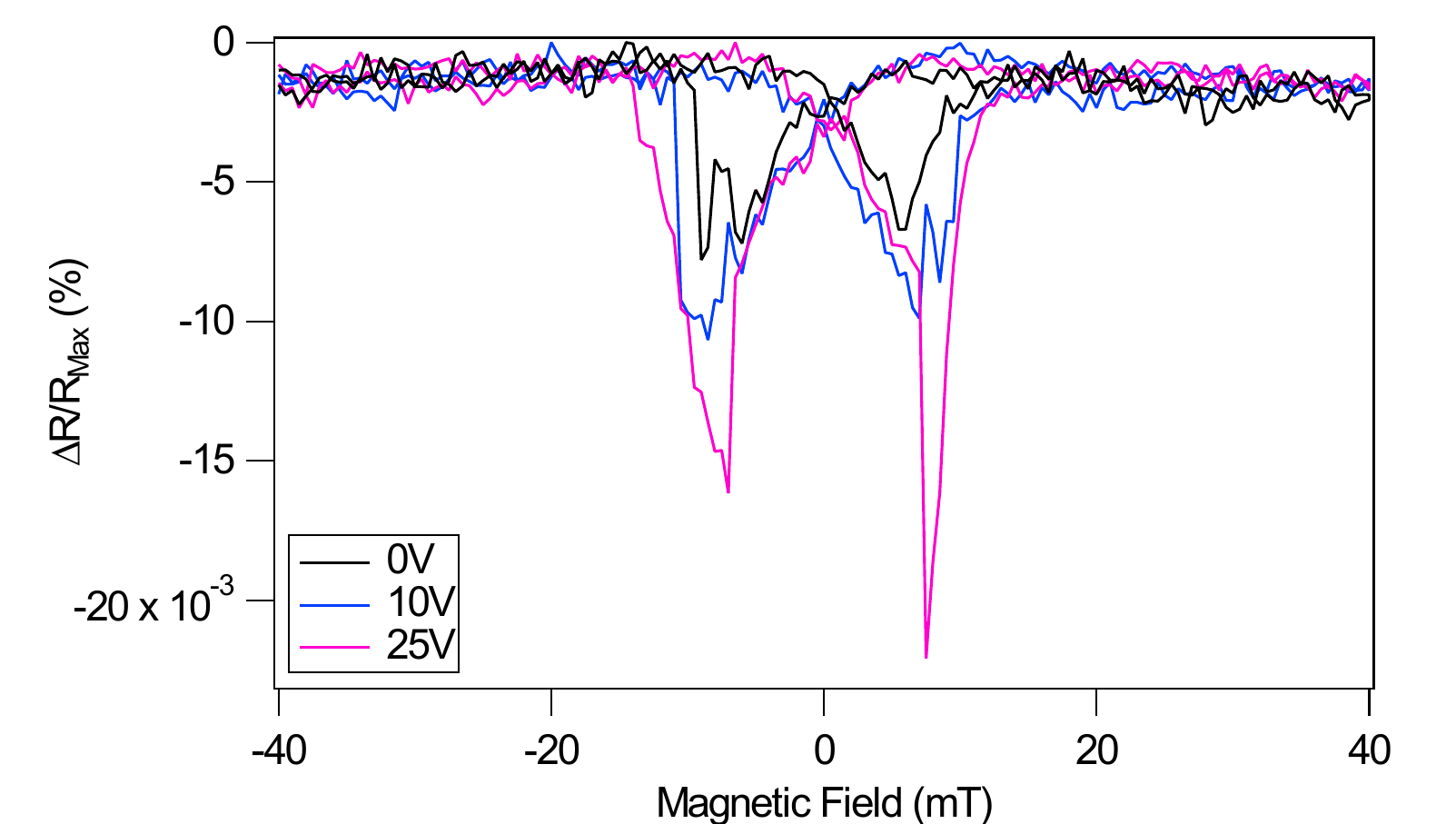}\label{fig:DevcomsolB}}\hfill
\subfloat{\xincludegraphics[width=0.44\textwidth,label=(d)]{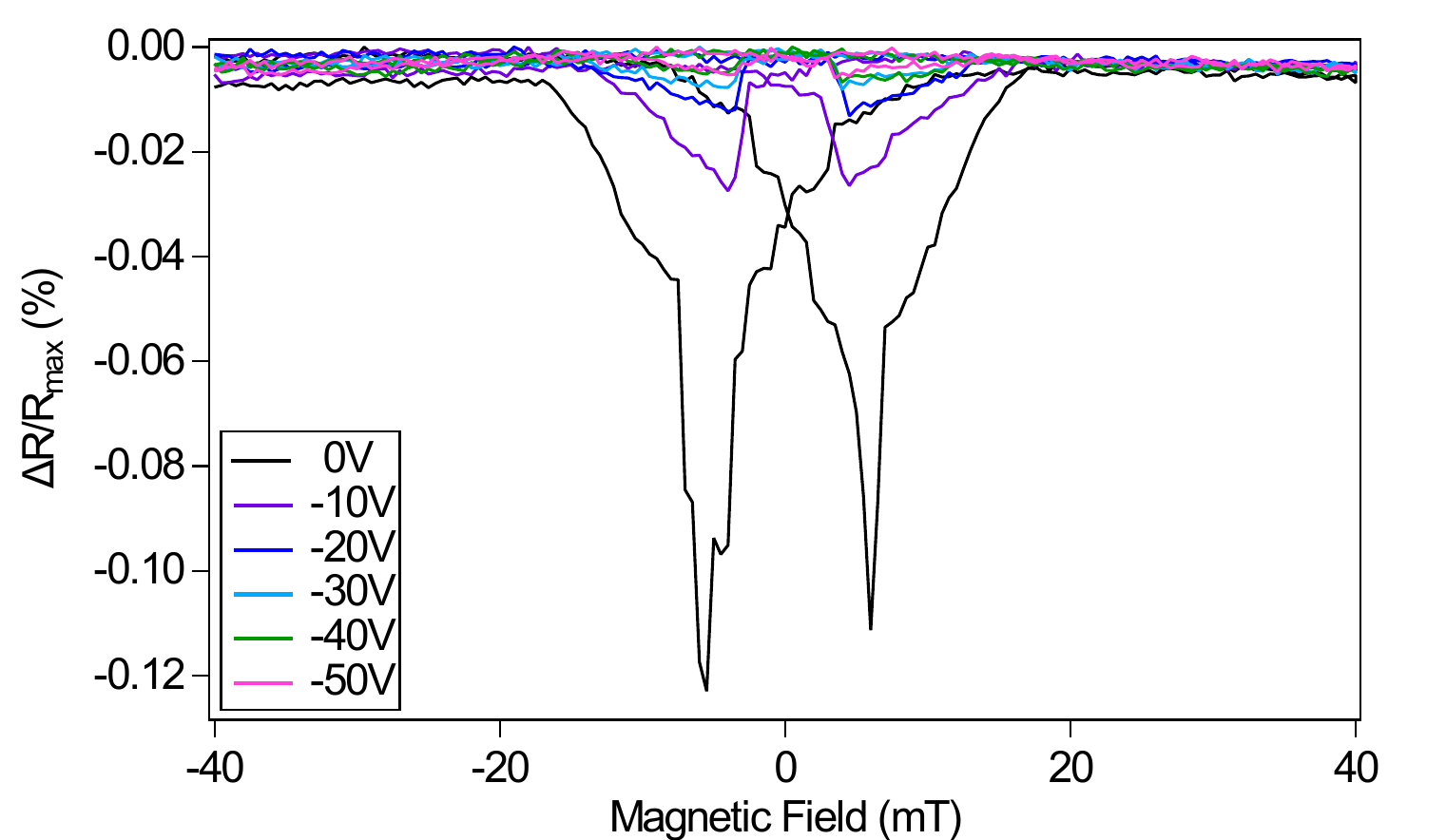}\label{fig:upsweepDevB}}\hspace{0.1cm}
\subfloat{\xincludegraphics[width=0.44\textwidth,label=(h)]{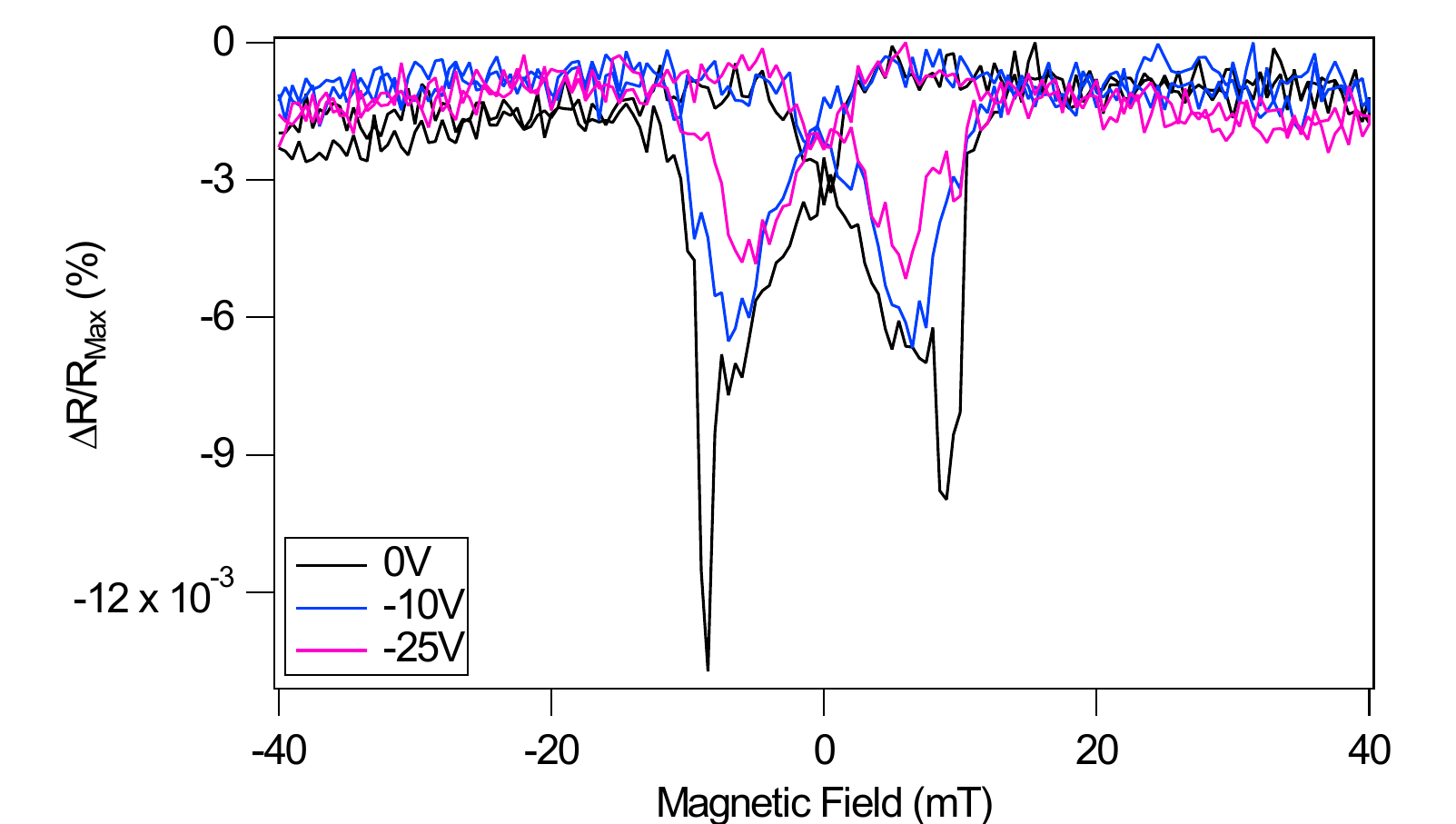}\label{fig:downsweepDevB}}\hfill
\caption{Magnetoresistance sweeps of Fe$_{0.75}$Co$_{0.25}$ taken at different applied voltages for device A (a, b, e, f) and device B (c, d, f, h). Microscope images of device A (a) and B (e) are displayed, with a circuit diagram added. Magnetic field in the MR sweeps was applied in-plane in the direction shown in (a) and (e) on the devices. COMSOL simulations are shown in (b) and (d) for the electric field being applied to the devices. Measurements were taken using a Wheatstone bridge to better measure the change in resistance. Measurements were taken in steps, with the 0\,V gating measurement depending greatly on the previous scan. First, the gate voltage was swept in the positive direction in steps of 10\,V to 50\,V, with the final measurement being taken after going to 100\,V and back to 50\,V, called the ‘upsweep’, shown in (e) and (f). The same procedure was applied in the negative direction, called the ‘downsweep’, shown in (g) and (h). The scans taken at 0\,V with different history are shown in SI for comparison. 
  }
  \label{fig:LinearAMR}
\end{figure*}

\begin{figure*}
{\subfloat{\xincludegraphics[width=0.45\textwidth,label=(a)]{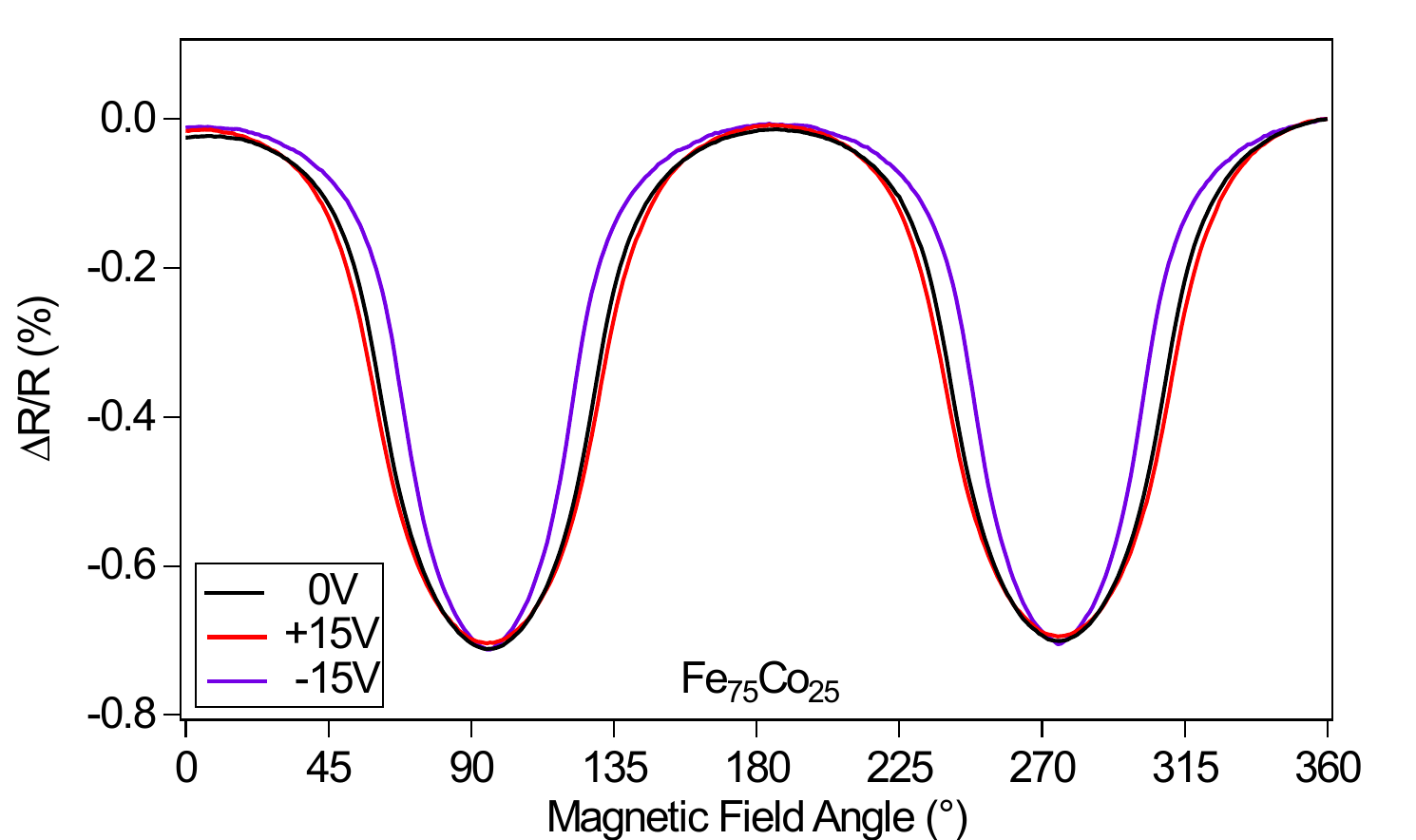}\label{fig:ADMRa}}\hfill}
{\subfloat{\xincludegraphics[width=0.45\textwidth,label=(b)]{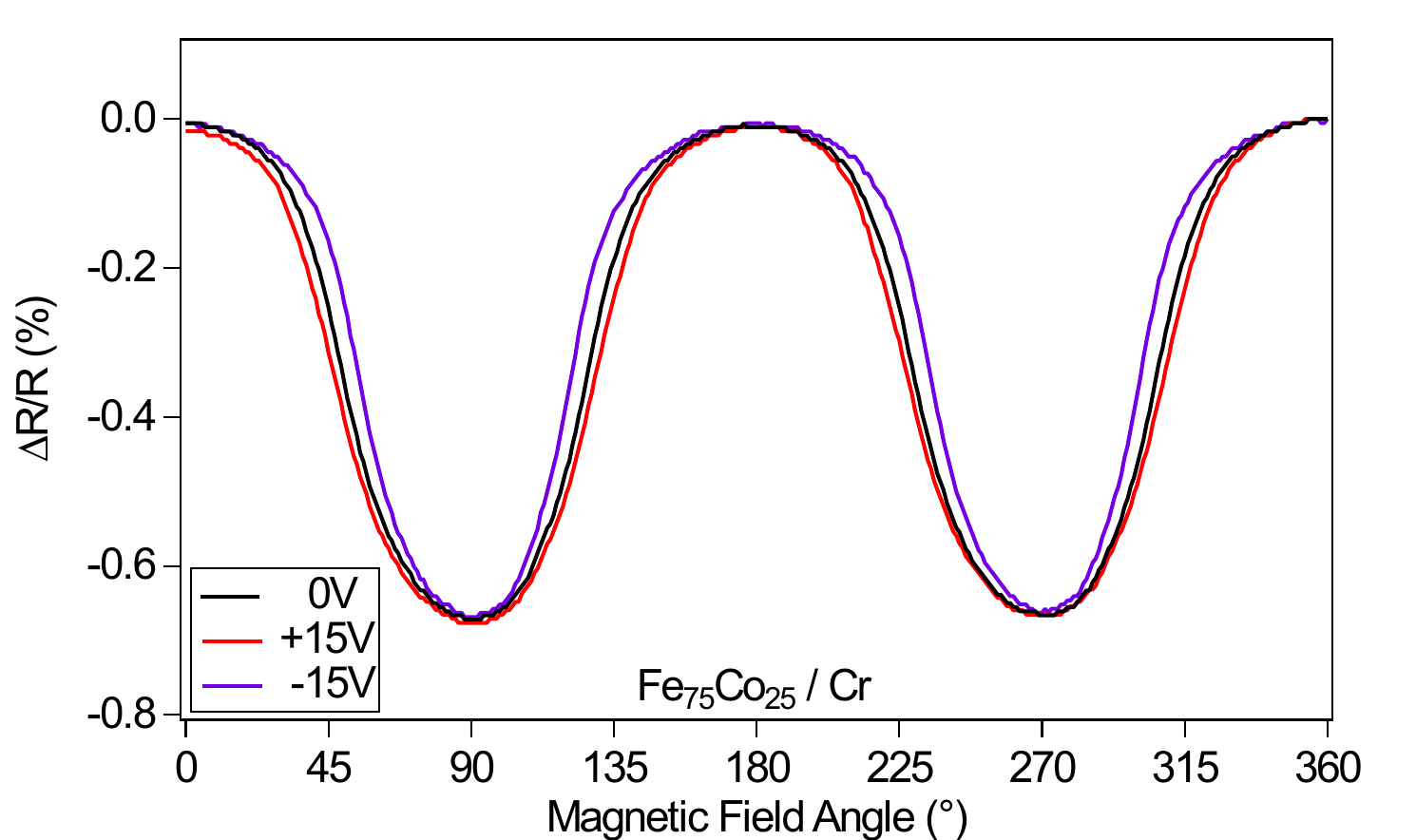}\label{fig:ADMRb}}\hfill}
\caption{Angle-dependent magnetoresistance for representative side-gate voltages in (a) Fe$_{0.75}$Co$_{0.25}$/PMN-PT (device B) with a magnetic field magnitude of 100\,mT and (b) Fe$_{0.75}$Co$_{0.25}$/Cr/PMN-PT (device C) with a magnetic field magnitude of 100\,mT. 
  }
  \label{fig:ADMR}
\end{figure*}

\section{Magnetoresistance and Magnetic Anisotropy}
\subsection{Electric Field Control of Magnetoresistance}
Magnetotransport measurements are used to investigate the magnetic properties and magneto-electric coupling of the Fe$_{0.75}$Co$_{0.25}$/PMN-PT heterostructures. 
Using a vector electromagnet to apply an in-plane external magnetic field, magnetoresistance (MR) is measured on devices patterned by a single step of photolithography followed by Ar ion milling. 
Figures~\ref{fig:DevDiaA} and \ref{fig:DevDiaB} show two device geometries used in this study.
In the first geometry, the device is patterned to have a conducting channel that is 20\,$\mu$m wide, 240\,$\mu$m long, with four electrodes and a single side gate to the left that is 20\,$\mu$m away from the channel (see Fig.~\ref{fig:DevDiaA}). In the second geometry, the device has a channel length of 700\,$\mu$m and width of 5\,$\mu$m. The distance to side gates is 10\,$\mu$m (see Fig.~\ref{fig:DevDiaB}).

For MR measurements in the first geometry (device A), an external magnetic field is applied along the device channel and ramped between $-40$\,mT and $40$\,mT while two-probe resistance measurements are performed using the electrodes shown in Fig.~\ref{fig:DevDiaA}.
During MR measurements, a side gate voltage is applied with reference to the device channel. 
Since the current applied to the channel is less than 100\,$\mu$A, there is $\sim$10\,mV voltage difference at the opposite ends (for 630\,$\Omega$).
This is much smaller compared to the side gate voltage applied, making the gating uniform along the device length. 
The gate creates an electric field in the substrate as shown by a COMSOL simulation. 
Figure~\ref{fig:DevcomsolA} shows a cross-section of the device channel on top of the PMN-PT substrate, and the side gate is to the left. 
The electric field generated by a positive voltage on the side gate is indicated by the arrows, where the color is the calculated electric field magnitude by position.  
The electric field underneath the device channel has both in-plane and out-of-plane components, and shows the magnitude of electric field applied underneath the device channel is only around 8\,kV/cm.

Figure~\ref{fig:upsweepDevA} shows MR hysteresis loops measured on device A for a series of positive side gate voltages. 
Here, the side gate voltage is increased in steps of 10\,V from 0\,V to 50\,V and an MR hysteresis loop is measured at each step. 
The ``upsweep'' of voltage causes a change in both the shape as well as the size of the MR curve, with the resistance drop becoming wider and larger with increasing voltage. 
The upsweep is followed by a ``downsweep'' to investigate negative gate voltages, as shown in Figure~\ref{fig:downsweepDevA}. 
After returning the gate back to 0\,V, the voltage is stepped from 0 to $-50$\,V in steps of $-10$\,V. In this case, the drops in resistance become narrower and smaller as the gate voltage becomes more negative. 
Together, the two voltage sweeps reveal a consistent trend where more positive voltages produce a larger magnitude of MR while more negative voltages produce a smaller magnitude of MR. 
Both the changes in the width and magnitude of the MR loops with gate voltage suggest a change in the magnetic anisotropy of the Fe$_{0.75}$Co$_{0.25}$ film, which is studied in detail later.

To better control the direction of the applied electric field, we employ a dual-gated geometry with a longer device channel, as shown in Fig.~\ref{fig:DevDiaB}.
Device B fabricated in this geometry has an Fe$_{0.75}$Co$_{0.25}$ thickness of 10\,nm, and 
the channel is along the $[110]_{PMN-PT}\parallel[100]_{FeCo}$ in-plane direction (i.e.~easy axis of the cubic anisotropy).
The COMSOL simulation of the electric field is shown in Figure \ref{fig:DevcomsolB}. Applying the same positive voltage to the two side gates (outside of the image) produces a field distribution with both in-plane and out-of-plane components of the electric field.
 
Next, we experimentally investigate the side-gate control of the MR hysteresis loop with this device structure. 
As shown in Figure \ref{fig:DevDiaB}, the same voltage is applied to the two side gates with respect to the device channel.
The upsweep voltage scan is taken from 0\,V to 25\,V in steps of 5\,V, with steps 10\,V and 25\,V shown in Figure \ref{fig:upsweepDevB}. The other steps are removed for better clarity, and are shown in SI. Similarly, the steps $-10$\,V and $-25$\,V of the downsweep voltage scan are displayed in Figure \ref{fig:downsweepDevB}, with intermediate steps shown in SI. 
In this new device geometry, we still see the same effects as observed in device A, although the magnitudes of the effects are smaller. 

The shapes of the MR hysteresis loops suggest that the resistance changes are due to anisotropic MR where the resistance with magnetization $\mathbf{M}$ parallel to the current ($R_{\parallel}$) differs from the resistance with $\mathbf{M}$ perpendicular to the current ($R_{\perp}$). Since perpendicular domains typically form during magnetization reversal, this causes the overall resistance to change from $R_\parallel$ toward $R_\perp$. If the observed MR does indeed originate from anisotropic MR, its dependence on the side gate voltage could be due to changes in the transport coefficients (i.e.~$R_\parallel$, $R_\perp$) and/or changes in the magnetic anisotropy. We investigate these possibilities next.

Fig~\ref{fig:LinearAMR} is to describe an intuitive understanding that the anisotropy is changing, decisively shown with device A. 
Device B allows for larger electric field application with smaller voltages and better saturation of the PMN-PT underneath the device channel. 
For a more quantitative analysis regarding the change in magnetic anisotropy of the Fe$_{0.75}$Co$_{0.25}$, we measure angle-dependent MR (ADMR) on device B. 

\subsection{Angle-Dependent Magnetoresistance}

Angle-dependent magnetoresistance (ADMR) scans are useful for confirming the presence of anisotropic MR, quantifying the transport coefficients $R_\parallel$ and $R_\perp$, and quantifying the magnetic anisotropy.
In these ADMR scans, the magnitude of the magnetic field is kept constant, while the angle, $\theta_H$, between magnetic field and channel axis is varied. 
Consistent with the anisotropic magnetoresistance origin, we observe maximum resistance when $\theta_H=0\degree$ and minimum when $\theta_H=90\degree$, as shown in Fig~\ref{fig:ADMR}. 
When applying the rotating magnetic field, the magnetic moment of the material follows, but lags behind, due to the magnetic anisotropy. 

The resistance curve is related to the change in the magnetic moment via the equation
\begin{equation}
    R(\theta_H) = R_{\perp}+(R_{\parallel}-R_{\perp})\cos^2(\theta_M),
    \label{eq:AMR}
\end{equation}
where $R$ is the measured resistance, and $\theta_M$ is the angle of magnetic moment measured from the channel axis.  
Analyzing the deviation of the ADMR curve from the perfect $\cos^2{\theta_H}$, one can extract anisotropy constants of the material.

As seen in Fig.~\ref{fig:ADMRa}, the gating induces a change in the shape of the ADMR curve for the Fe$_{0.75}$Co$_{0.25}$. 
The applied negative voltage increases the anisotropy, as the curve deviates more from $\cos^2{\theta_H}$, while the positive voltage has the opposite effect.   
The observed effect is an odd function of applied electric field, which poses the question of the origin of the magneto-electric coupling. 
To better investigate this, another film of Fe$_{0.75}$Co$_{0.25}$ is grown on PMN-PT with a 7\,nm Cr interlayer 
to eliminate purely interfacial coupling (i.e.~Fe$_{0.75}$Co$_{0.25}$/Cr/PMN-PT). 
A device (device C) is patterned from this film in the dual-gated geometry (Fig.~\ref{fig:DevDiaB}) and the ADMR is measured, as shown in Fig~\ref{fig:ADMRb}.
We observe similar gating-induced changes in ADMR curves with and without the Cr buffer.
To investigate, we proceed to a quantitative analysis of the ADMR data. 

\subsection{Magnetic Anisotropy Analysis}

We analyze the ADMR scans to quantify the magnetic anisotropy following the procedure in Hu \textit{et al}~\cite{hu_determination_2015}. 
Because the magnetization remains predominantly single domain, we use the Stoner-Wohlfarth model \cite{stoner-wohlfarth-model}. The energy is given by
\begin{equation}
\begin{split}
E = K_{u}&\sin^{2}(\theta_{M})
+ K_{c}\sin^2(\theta_{M})\cos^2(\theta_{M}) \\
&-\mu_{0}M_s H \cos(\theta_M-\theta_H),
\end{split}
\label{eq:anisotropy}
\end{equation}
\newline where $K_{u}$ is an in-plane uniaxial anisotropy, $K_{c}$ is the cubic magnetocrystalline anisotropy, $\mu_0$ is the permeability of free space, $M_s$ is the saturation magnetization, and $H$ is the magnitude of the applied magnetic field. Minimizing the energy by setting $\partial{E} / \partial{\theta_M} = 0$ yields the equation for magnetic torque,
\begin{equation}
\begin{split}
    \mu_{0}M_s H \sin(\theta_H-\theta_M) = & \, K_{u}\sin(2\,\theta_{M})\\
    & + \frac{1}{4}K_{c}\sin(4\,\theta_{M})
    \label{eq:torque}
\end{split}
\end{equation}

The values for $\theta_M$ are determined from the ADMR data $R(\theta_H)$ (Figure \ref{fig:ADMR}) by inverting Eq.~(\ref{eq:AMR}) to get $\theta_M$ as a function $\theta_H$, i.e. $\theta_M = \cos^{-1}\left(\pm \sqrt{\frac{R(\theta_H)-R_\perp}{R_\parallel-R_\perp}}\right)$. 
As the inversion formula is multivalued, the value of $\theta_M$ closest to $\theta_H$ should be selected.
This makes Eq.~(\ref{eq:torque}) to be a function of $\theta_H$ alone. The magnetic anisotropy coefficients are determined by plotting the left-hand side (LHS) of Eq.~(\ref{eq:torque}) and fitting with the right-hand side (RHS) of Eq.~(\ref{eq:torque}) using $K_u$ and $K_c$ as fitting parameters.

\begin{figure}[h!]
\subfloat{\xincludegraphics[width=0.45\textwidth]
{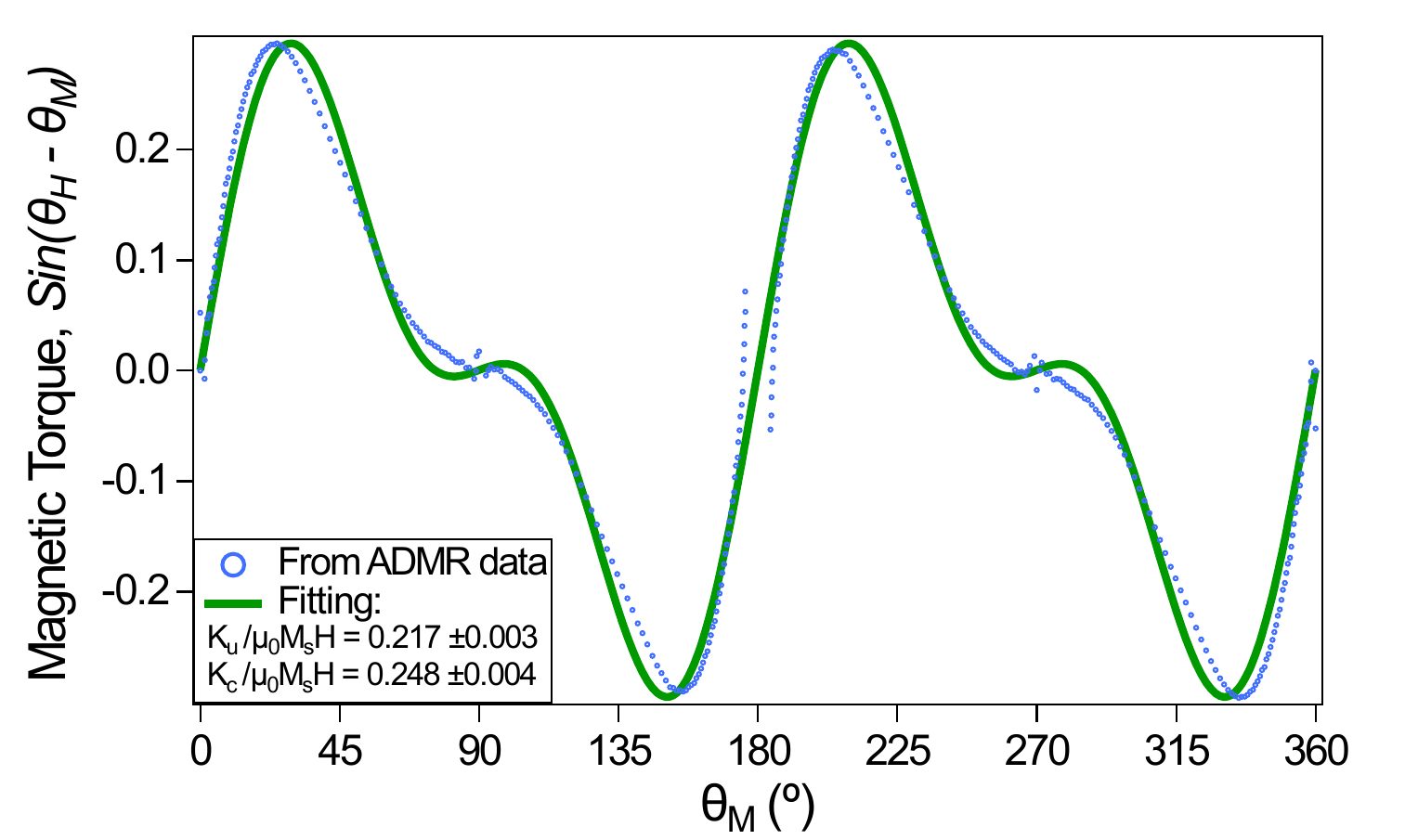}\label{fig:fit_FeCo}}\hfill

\vspace{-0.01\textwidth}
\caption{Magnetic torque fitting to extract anisotropy constants for Fe$_{0.75}$Co$_{0.25}$ (device B). 
  }
  \label{fig:fit}
\end{figure}

\begin{figure}
\subfloat{\xincludegraphics[width=0.40\textwidth,label=(a)]{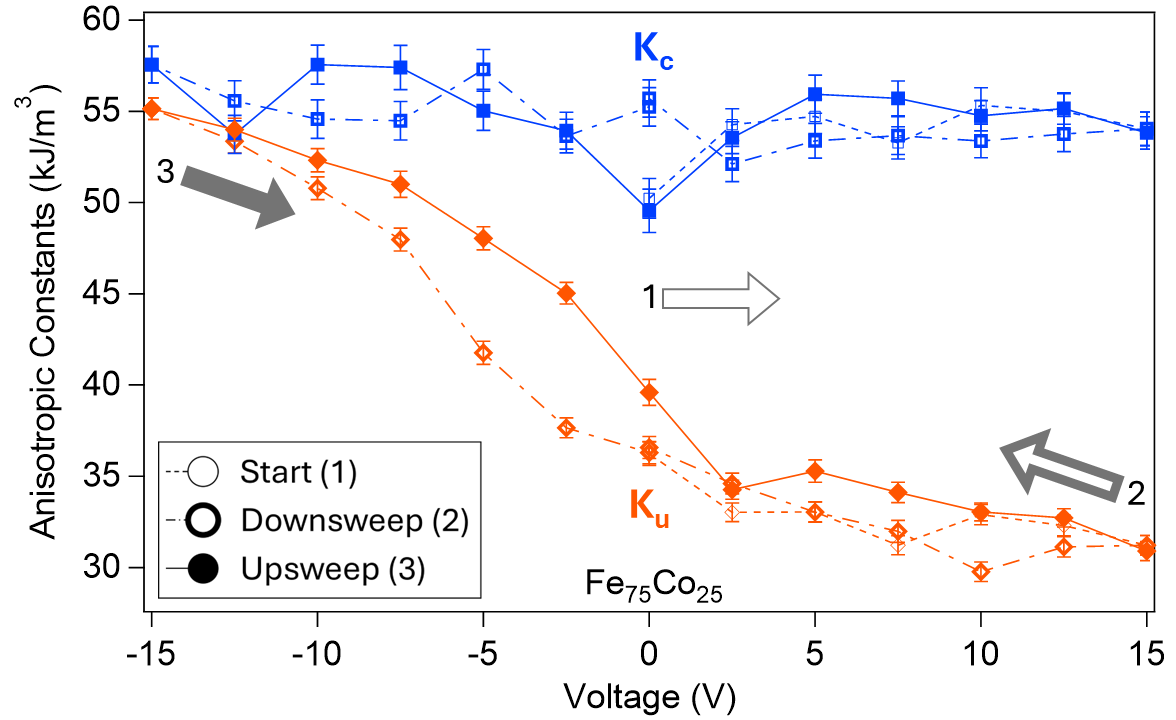}\label{fig:K_FeCo}}\hfill
\subfloat{\xincludegraphics[width=0.40\textwidth,label=(b)] {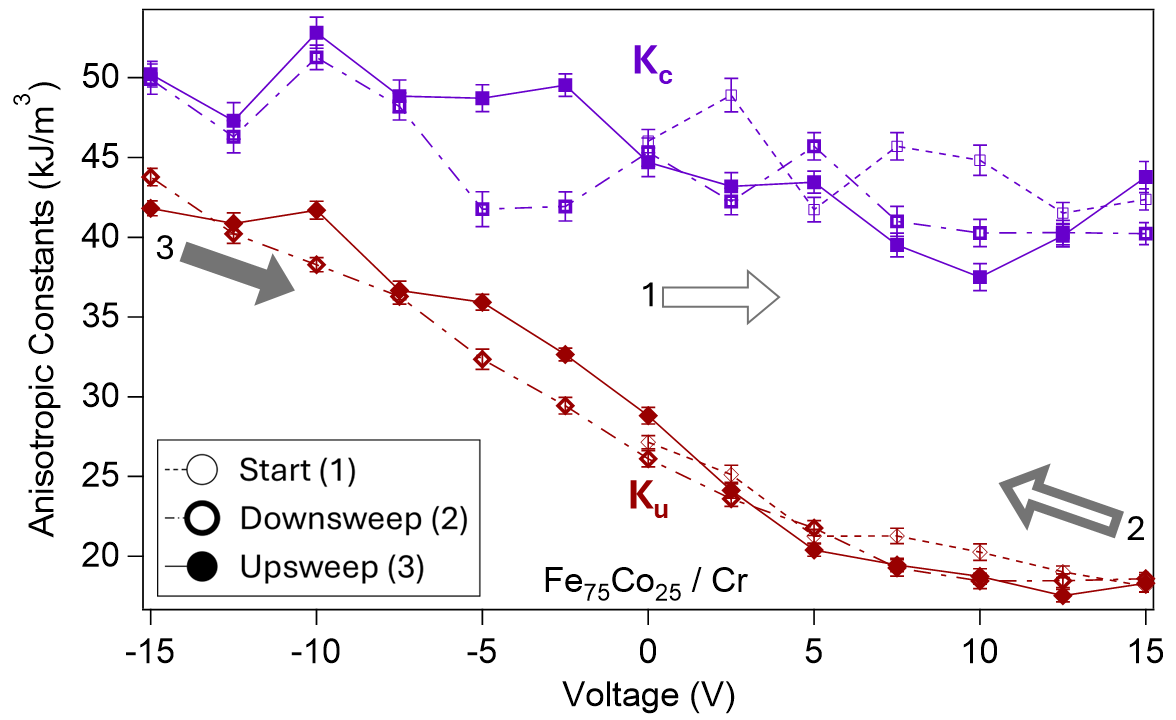}\label{fig:K_FeCoCr}}\hfill
\subfloat{\xincludegraphics[width=0.40\textwidth,label=(c)]{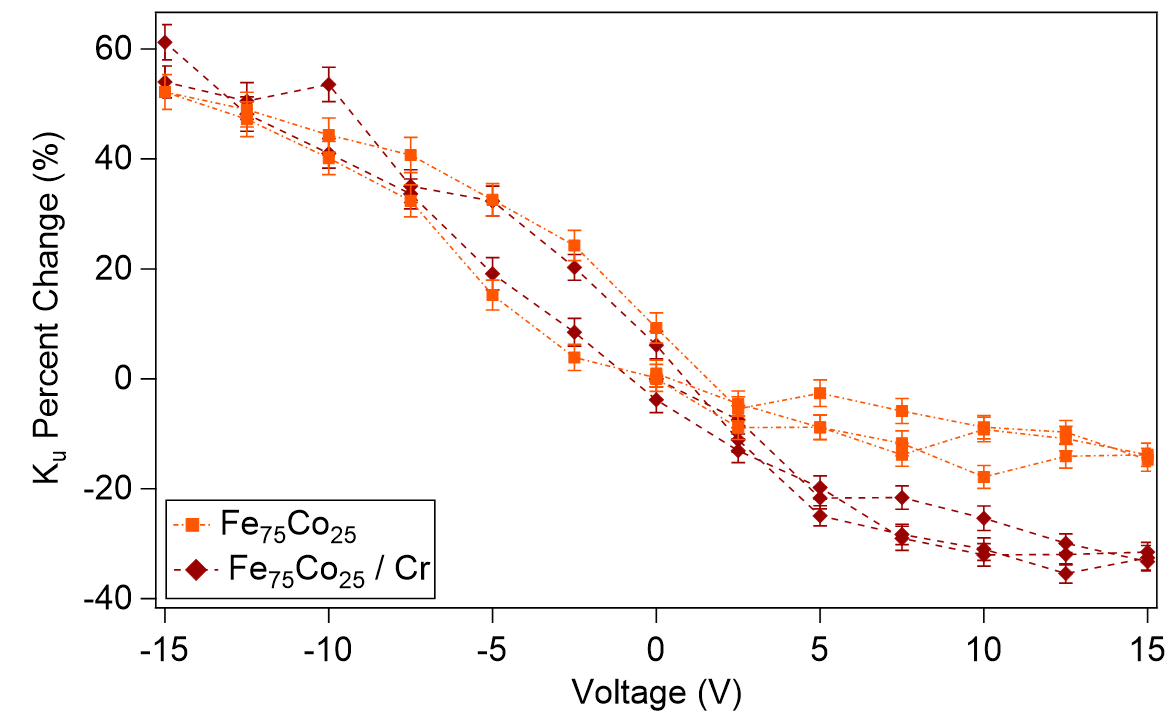}\label{fig:Ku_compare}}\hfill
\caption{Uniaxial and cubic magnetocrystalline anisotropy coefficients as a function of applied electric field (side-gate voltage). (a) Results for device B, Fe$_{0.75}$Co$_{ 0.25}$/PMN-PT using $\mu_0 H = 100$\,mT for the ADMR scans. (b) Results for device C, Fe$_{0.75}$Co$_{ 0.25}$/Cr/PMN-PT using $\mu_0 H = 100$\,mT for the ADMR scans. (c) Comparing the uniaxial anisotropies of devices B and C, plotted as a percent change from zero applied field.
  }
  \label{fig:K_vs_E}
\end{figure}

Figure~\ref{fig:fit} illustrates this procedure applied to the data in Figure~\ref{fig:ADMRa}. The magnetic torque calculated from the ADMR data (LHS) is plotted as open circles and the fitting by $K_u$ and $K_c$ (RHS) is plotted as the solid line. 
The fitting with $K_u$ and $K_c$ terms account well for the observed data.
The process of analyzing the raw data is discussed further in the Supplemental Material (SM). \cite{SupplMat}

To understand how the magnetic anisotropy varies with applied electric field, we measured ADMR scans (with $\mu_0 H = 100$\,mT) as a function of side-gate voltage and extracted $K_c$ and $K_u$.
Figure~\ref{fig:K_FeCo} shows the results for device B (Fe$_{0.75}$Co$_{0.25}$/PMN-PT) and Figure~\ref{fig:K_FeCoCr} shows the results for device C (Fe$_{0.75}$Co$_{0.25}$/Cr/PMN-PT). 
In both cases, the side-gate voltage starts at 0\,V, sweeps up to $+15$\,V (18\,kV/cm$^2$) with a step size of 2.5\,V, sweeps down to $-15$\,V, then back up to $+15$\,V. 
What is most notable in both samples is that $K_u$ shows a strong variation with applied electric field, but $K_c$ does not. 
Figure~\ref{fig:Ku_compare} compares the variation of $K_u$ for the two samples by replotting them as a percentage change from their zero-electric-field values. 
These show similar electric-field dependence of the uniaxial anisotropy for Fe$_{0.75}$Co$_{0.25}$/PMN-PT and Fe$_{0.75}$Co$_{0.25}$/Cr/PMN-PT, whose significance is discussed in the next section.

\section{Discussion}

The experimental results show that the electrical control of the MR hysteresis loops is due to a combination of anisotropic MR and electric-field-induced changes to the in-plane uniaxial magnetic anisotropy $K_u$. 
Knowing that the electrical control of the MR hysteresis loops is related to the magnetic anisotropy, the smaller effects in device B compared to device A (Fig.~\ref{fig:LinearAMR}) could be understood as follows. 
Device B has a narrower and longer device channel than device A. With 5\,nm versus 20\,nm wide, the geometry of device B enhances the magnetic shape anisotropy, pinning the magnetization along current direction. This makes it more difficult for perpendicular (in-plane) domains to form. In addition, if the magnetization reversal is due to domain wall motion, these perpendicular domains are likely to constitute a smaller fraction of the domains in the channel, thus contributing less to the overall channel resistance. These two effects will reduce the changes to the MR hysteresis loops. For purposes of applications, the length-to-width aspect ratio should be designed to maximize the electric field effects.

We now turn our attention to the magneto-electric coupling mechanism for the observed electric field control of $K_u$.
Comparing Fe$_{0.75}$Co$_{0.25}$/PMN-PT (device B) and Fe$_{0.75}$Co$_{0.25}$/Cr/PMN-PT (device C) shows that both exhibit similar variations of $K_u$ as a function of gate voltage.
This observation provides support that the variations in $K_u$ originate from a strain-mediated effect (Figure~\ref{fig:ME_strain}) as opposed to an interfacial electric charge effect (Figure~\ref{fig:ME_charge}), such as that observed at Fe/MgO interfaces~\cite{maruyama_large_2009}. In those studies, electric fields applied across Fe/MgO/metal junctions produced changes in the surface perpendicular magnetic anisotropy. 
In our studies, we insert a Cr interlayer to distinguish between the strain vs.~charge effect. If the interface charge effect were the most important, the insertion of the Cr interlayer would move the interface charge layer away from the Fe$_{0.75}$Co$_{0.25}$ layer (i.e.~to the Cr/PMN-PT interface), and thereby remove or greatly reduce the interface charge effect on the magnetic anisotropy. 
Since the insertion of the 7\,nm Cr interlayer has little effect on the gate dependence of $K_u$, this argues against the charge effect as the origin of the anisotropy change.
However, it has been argued previously that the charge-mediated effect could propagate further than~2\,{\AA} through a magnon-assisted magnetoelectric coupling at a FM/FE interface~\cite{jia_mechanism_2014,zhou_long-range_2018}. This is unlikely to be relevant in our study, due to the Cr being non-magnetic at room temperature. 

In contrast, a strain-mediated effect is better able to explain the experimental observations. Whereas electric charge effects are rapidly screened in metals, the strain effects could propagate over larger distances. This provides a natural reason why the insertion of a Cr interlayer does not substantially reduce the gate-control of $K_u$. 
However, it is worth noting that the changes in uniaxial anisotropy are odd across zero rather than even. In terms of symmetry, a charge-mediated effect is odd in the electric field, whereas a strain-mediated effect is typically even \cite{zhou_long-range_2018,jia_mechanism_2014}. 
However, it has been shown that a strain-mediated effect in PMN-PT(001) could exhibit an odd-like symmetry as opposed to even \cite{ba_electric-field_2021,zhang_electric-field_2024}. 
In our case, an odd-like symmetry could occur if the applied electric field is insufficient to switch the polarization of PMN-PT. 
In our side gate geometry, the largest electric fields are restricted to the surface regions (see Fig.~\ref{fig:DevcomsolA} and \ref{fig:DevcomsolB}), so larger gate voltages may need to be applied to induce ferroelectric switching.

Furthermore, a strain-mediated effect would provide a natural explanation for the variations in the \textit{in-plane} uniaxial anisotropy. For the side gate geometry, the applied voltage creates both in-plane and out-of-plane components of the electric field (Figures \ref{fig:DevcomsolA} and \ref{fig:DevcomsolB}). The in-plane component, which breaks the symmetry of the two in-plane directions, exists not only underneath the edge regions the Fe$_{0.75}$Co$_{0.25}$ channel, but also dominates in the regions between the Fe$_{0.75}$Co$_{0.25}$ channel and the side gates. The piezoelectric effect will produce an in-plane uniaxial strain in these regions that will also extend into the Fe$_{0.75}$Co$_{0.25}$ channel, thereby modifying the in-plane uniaxial magnetic anisotropy through the magnetoelastic coupling. On the other hand, a charge effect would predominantly alter the out-of-plane perpendicular magnetic anisotropy as opposed to an in-plane uniaxial anisotropy.


To further evaluate the strain-mediated effect, we analyze the variation of $K_u$ in conjunction with COMSOL simulations that estimate the strain on the Fe$_{0.75}$Co$_{0.25}$ channel. First, the shape of the $K_u$ vs.~gate voltage follows a quasi-linear shape on both the up sweep and down sweep between $+15$\,V and $-15$\,V. This is in contrast to hysteretic butterfly-shaped curves of strain vs.~applied electric field observed in bottom-gated ferromagnet/PMN-PT(001) samples \cite{schoffmann_strain_2022}. The main difference is that the butterfly shape occurs when a full ferroelectric hysteresis loop is traversed, while a quasi-linear shape occurs when a minor loop is traversed (i.e.~the applied electric field remains below the coercive field). From the data in Figure \ref{fig:K_FeCo}, we quantify the variation of $K_u$ with gate voltage by its average slope between $-15$\,V and $+15$\,V: $\frac{\Delta K_u}{\Delta V_g} = \left( \frac{3.1 \times 10^4\,\text{J/m}^3 - 5.5 \times 10^4\,\text{ J/m}^3}{30\,\text{V}} \right) = -800\,\frac{\text{J}}{\text{m}^3 \text{V}}$. Because the sign of the slope depends on the remnant polarization state which is unknown, we can only consider the magnitude of the slope, $\left| \frac{\Delta K_u}{\Delta V_g}  \right| = 800\,\frac{\text{J}}{\text{m}^3 \text{V}}$.

Details of the COMSOL simulations of the strain and analysis of the magnetoelastic anisotropy are provided in SM section S3, and only the key results are highlighted in the following.
With $+15$ V applied to the side gates, we find that the in-plane strains along the channel ($\varepsilon'_{11} = -0.19\times10^{-4}$) and perpendicular to the channel towards the side gates ($\varepsilon'_{22} = -3.69\times10^{-4}$) are compressive and the strain perpendicular to the substrate surface ($\varepsilon'_{33}$) is tensile (note that $\varepsilon'_{ij}$ is the strain in the coordinates of the Fe$_{0.75}$Co$_{0.25}$ cubic unit cell which is aligned with the device channel; the PMN-PT unit cell differs by a $45\degree$ in-plane rotation). 
As derived in SM section S3, this strain produces a magnetoelastic contribution to the in-plane uniaxial anisotropy given by $K_u^{me} = B_{1}(\varepsilon'_{22}-\varepsilon'_{11})$, where the magnetoelastic constant $B_1$ is defined in refs.~\cite{becker_ferromagnetismus_1939,wedler_magnetoelastic_2000,Sander2021}.
Using the experimental value of $\left| \frac{\Delta K_u}{\Delta V_g}  \right| = 800\,\frac{\text{J}}{\text{m}^3 \text{V}}$ and the strain calculated in COMSOL, we determine the magnitude of the magnetoelastic constant to be
$\left| B_1 \right| = \left| \frac{K_u^{me}}{\varepsilon'_{22}-\varepsilon'_{11}} \right| = \left| \frac{(\Delta K_u/\Delta V_g)*(V_g)}{\varepsilon'_{22}-\varepsilon'_{11}} \right| = \left| \frac{(800\frac{\text{J}}{\text{m}^3 \text{V}})*(15\,\text{V})}{-3.69\times10^{-4}+0.19\times10^{-4} \frac{J}{m^3}} \right| = 35$\,MJ/m$^3$. Because the sign of $B_1$ is negative for FeCo \cite{serizawa20191905R003}, our anisotropy data and strain simulations yield a value of $B_1 = -35$\,MJ/m$^3$.

Since we were unable to find an independent measurement of $B_1$ in single-crystalline Fe$_{0.75}$Co$_{0.25}$, we estimated its value through $B_1=-\frac{3}{2}\lambda_{100}(c_{11}-c_{12})$ \cite{Sander2021}, recent measurements of the magnetostriction constant $\lambda_{100}$ of single-crystalline Fe$_{0.70}$Co$_{0.30}$ films \cite{serizawa20191905R003}, and experimental and theoretical studies of the elastic constants $c_{11}$ and $c_{12}$ in FeCo alloys \cite{adams_elastic_2006,hossain_systematic_2020} (details in SM section S3). This yields an estimate of $B_1$ between $-34$ MJ/m$^3$ and $-53$ MJ/m$^3$, which shows that the value of $-35$ MJ/m$^3$ extracted from our experimental data and COMSOL simulations is quite reasonable. Thus, this analysis provides additional support that the magnetoelectric coupling originates from a strain-mediated effect.

\section{Conclusion}

A side-gated device geometry was designed and implemented to study magnetoelectric coupling in a system of Fe$_{0.75}$Co$_{0.25}$ and PMN-PT. 
We found that the magnetoresistance hysteresis loops could be controlled by the side-gate voltage. 
To investigate the origin of this effect, angle-dependent magnetoresistance measurements were taken, and the dominant effect was shown to be electrical control of the magnetic anisotropy as opposed to the transport coefficients.
It was discovered that there was very little change in the cubic magnetocrystalline anisotropy, but as large as 60\% change in the in-plane uniaxial anisotropy. 
This effect was observed with similar magnitude even when a 7\,nm Cr interlayer was inserted between the Fe$_{0.75}$Co$_{0.25}$ and the PMN-PT substrate. 
This argues that the magneto-electric coupling for this system is strain-mediated rather than interfacial-charge-mediated.
The electrical field control of magnetic anisotropy with side-gate geometry demonstrated in this work presents new opportunities for the electrical generation of magnons and electrical tuning of their propagation in surface acoustic wave applications.

\section*{Acknowledgments}

This work received primary support from the Defense Associated Graduate Student Innovators (DAGSI) Award No. RX-22, partial support from the Air Force Office of Scientific Research (AFOSR) under award number FA955023RXCOR001, and partial support from the Center for Emergent Materials, an NSF MRSEC, under award number DMR-2011876.

\bibliography{refs.bib}

\end{document}



\title{Supplemental Material:\\ Electrical Side-Gate Control of Anisotropic Magnetoresistance and Magnetic Anisotropy in a Composite Multiferroic}

\author{Katherine Johnson}
\email{robinson.1971@buckeyemail.osu.edu}
\affiliation{Department of Physics, The Ohio State University, Columbus, Ohio 43210, United States}

\author{Michael Newburger}
\affiliation{Materials and Manufacturing Directorate, Air Force Research Laboratory, Wright-Patterson Air Force Base, Ohio 45433, USA}
\author{Michael Page}
\affiliation{Materials and Manufacturing Directorate, Air Force Research Laboratory, Wright-Patterson Air Force Base, Ohio 45433, USA}
\author{Roland K. Kawakami}
\affiliation{Department of Physics, The Ohio State University, Columbus, Ohio 43210, United States}

\maketitle


\newpage
\section{\NoCaseChange{COMSOL Electrostatics Simulations}}
To determine the electric field created by the side gates, as shown in Figure 3 of the main text, COMSOL Multiphysics was utilized to simulate the electric field distribution within a specified domain. 
The simulation process began by selecting the ``Electric Fields'' module in a stationary study (no time dependence). 
A 2D geometry was created representing the physical systems  of both device A and device B, sliced perpendicular to the channel axis, with iron (mat4) used as the material for the electrodes (in lieu of Fe$_{0.75}$Co$_{0.25}$) and adjusted PZT(4) for the substrate. The measurements of the structures defined in the simulation is displayed in Figure \ref{fig:comsol-dev-diagram}.

\begin{figure}[h]
    \includegraphics[width=0.45\textwidth]{figures/Supplemental_Fig/New SI/deviceA_simul_def_draft1.png}
    \includegraphics[width=0.45\textwidth]{figures/Supplemental_Fig/New SI/deviceB_simul_def_draft1.png}
  \hfill
    \caption{Drawn structures in COMSOL used in the electrostatics simulations. Left is device A, right is device B. }
    \label{fig:comsol-dev-diagram}
\end{figure}

The geometry was meshed with the physics controlled sequence type with element size `fine' mesh to ensure accurate representation of the electric field gradients.
Boundary conditions were applied to the model, including either $+15$\,V or $-15$\,V on side gate surfaces, and ground on the current channel. 
The governing equations, such as Poisson's equation for electrostatics, were solved numerically using the finite element method (FEM) to determine the electric field distribution within the substrate.
After running the simulation, the electric field results were post-processed to visualize the field vectors and field strength, as seen in Figure 3.

\section{\NoCaseChange{Symmetrization of Angle-Dependent Magnetoresistance (ADMR) Data}}

\begin{figure}[h]
    \includegraphics[width=0.6\textwidth]{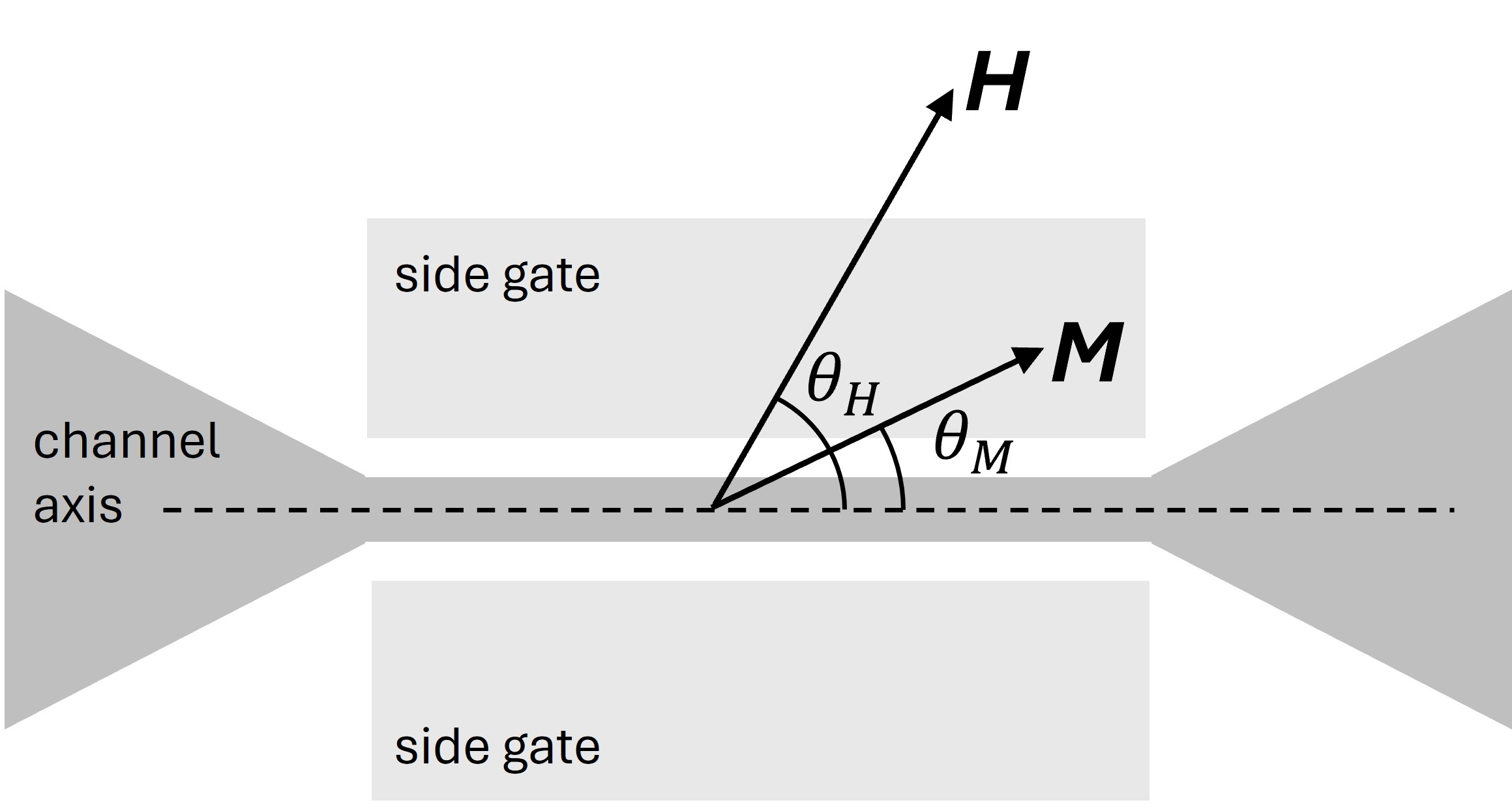}
  \hfill

    \caption{Diagram of the device geometry, applied magnetic field angle $\theta_H$, and magnetization angle $\theta_M$ }
    \label{fig:symmetrization}
\end{figure}

Due to a slight in-plane rotational misalignment when the sample is mounted, an offset angle is added to the raw ADMR data.  With this offset, the magnetic field angle $\theta_H$ and magnetization angle $\theta_M$ are referenced to the channel axis of the device as shown in Fig.~\ref{fig:symmetrization}.

\begin{figure}[h]
    \includegraphics[width=0.45\textwidth]{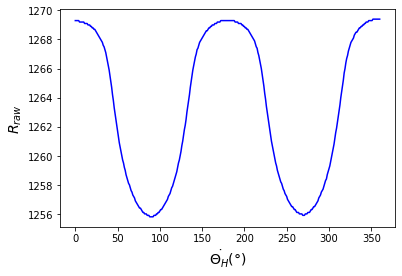}
    \includegraphics[width=0.45\textwidth]{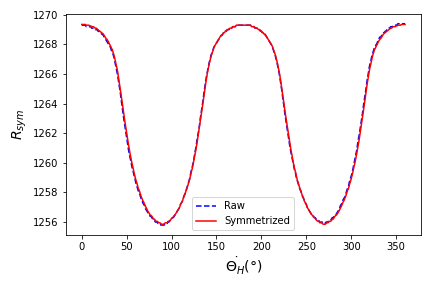}
  \hfill
    \caption{Comparison of the raw ADMR scan (left) and symmetrized ADMR scan (right). The symmetrized scan also shows the raw scan (in dashed blue) to help compare. }
    \label{fig:raw-vs-sym}
\end{figure}

We noticed a thermal drift in our data due to heating by the nearby magnet poles, which was greatly suppressed by water-cooling the electromagnet. However, a slight drift persisted, as shown in Figure \ref{fig:raw-vs-sym}. Because the device geometry, sample gating, and crystallographic axes all are symmetric with respect to reflection about the channel axis (see Fig.~\ref{fig:symmetrization}), we reduced the effect of thermal drift by symmetrizing the data about the channel axis using the equation
\begin{equation}
    R_{sym}(\theta_H) = \frac{R(\theta_H)+R(360\degree-\theta_H)}{2},
    \label{eq:AMR}
\end{equation}

When the magnetic anisotropy analysis was carried out for the original data and the symmetrized data (Figure \ref{fig:SI-KuKc}), similar trends were observed for the gate dependence of the cubic anisotropy $K_c$ and uniaxial anisotropy $K_u$. 
For the uniaxial anisotropy, the fitted $K_u$ values vs.~gate voltage extracted from the original ADMR scans and the symmetrized scans agree well, as seen in Figure \ref{fig:SI-KuKc}. 
For both the original and symmetrized scans, $K_u$ decreases substantially with increasing gate voltage. 
On the other hand, the extracted cubic anisotropy $K_c$ exhibits more noise as a function of gate voltage for the original data as compared to the symmetrized data. In addition, the numerical value of $K_c$ is slightly higher after symmetrization. However, neither set exhibits a strong or systematic variation of $K_c$.
Thus, symmetrizing the data does not affect the main conclusions that $K_u$ varies strongly with gating and $K_c$ exhibits little or no gate dependence.

\begin{figure}[h]
    \includegraphics[width=0.45\textwidth]{figures/Supplemental_Fig/sym vs no sym_Ku.pdf}
    \includegraphics[width=0.45\textwidth]{figures/Supplemental_Fig/sym vs no sym_Kc.pdf}
  \hfill
    \caption{Comparison of calculated anisotropy constants with and without symmetrization. Left: $K_u$ agrees within 2-bar error. Right: $K_c$ is numerically different, but it does not change qualitatively.}
    \label{fig:SI-KuKc}
\end{figure}

\section{\NoCaseChange{Piezoelectric Simulations and Magnetoelastic Anisotropy}}

\subsection {Piezoelectric Simulations}

\begin{figure}[h]
    \includegraphics[width=0.39\textwidth]{figures/Supplemental_Fig/New SI/piezosim_drawingdim_draft2.png}
    \includegraphics[width=0.53\textwidth]{figures/Supplemental_Fig/New SI/atomicstructure_figure_draft1.png}
  \hfill
    \caption{Diagram of the device drawn and used in simulation and a diagram of atomic structure of the materials with respect to the axis $xyz$ and $x'y'z'$. }
    \label{fig:piezosim_drawingdim}
\end{figure}

COMSOL simulations used the physics interfaces ``Solid Mechanics'' and ``Electrostatics,'' with the multiphysics interface ``Piezoelectric Effect.'' Electrostatics was used to define the voltage applied to the device to be $+15$\,V on both side gates while the channel was held at ground. A fixed force constraint was applied to the bottom of the PMN-PT block, and the ``strain-charge'' form was used. The study was performed in ``stationary'' mode. The simulations had the dimensions shown in Figure~\ref{fig:piezosim_drawingdim}, where the spacing is the same as the device. 
The geometry was meshed with the physics controlled sequence type with element size `extra fine' mesh.
The simulation used the finite element method (FEM) to determine the electric field distribution within the substrate and the strain from the applied electric field.
After running the simulation, the strain components were post-processed. 
The magnetic material of the device was chosen to be iron(mat4), and the PMN-PT material constants were taken from literature~\cite{PMN-PTconts}.  
The piezoelectric constants (Table~\ref{table:1}) are defined by the $xyz$ coordinates of the PMN-PT cubic-like unit cell. The device, however, is rotated in-plane by 45$\degree$ relative to the $xyz$ of the substrate because Fe$_{0.75}$Co$_{0.25}$ unit cell is 45$\degree$ with respect to substrate, with the magnetic easy axis along $x'$ and $y'$ seen in Figure~\ref{fig:piezosim_drawingdim}. 
This means that while the simulation must be done with respect to coordinates $xyz$, the analysis of strain needs to be in the rotated frame $x'y'z'$. 

\begin{table}[]
\begin{tabular}{cclclcl}
\hline
& \multicolumn{2}{c}{\cellcolor[HTML]{ECF4FF}d$_{15}$} 
& \multicolumn{2}{c}{\cellcolor[HTML]{ECF4FF}d$_{13}$} 
& \multicolumn{2}{c}{\cellcolor[HTML]{ECF4FF}d$_{33}$} 
\\ \cline{2-7} 
\multirow{-2}{*}{\begin{tabular}[c]{@{}c@{}}Piezoelectric Coupling Matrix, \\ Voigt notation (pC/N)\end{tabular}} & \multicolumn{2}{c}{1410}
& \multicolumn{2}{c}{-130}
& \multicolumn{2}{c}{307}
\\ \hline 
& \multicolumn{1}{c}{\cellcolor[HTML]{ECF4FF}s$_{11}$} & \multicolumn{1}{c}{\cellcolor[HTML]{ECF4FF}s$_{12}$} & \multicolumn{1}{c}{\cellcolor[HTML]{ECF4FF}s$_{13}$} & \multicolumn{1}{c}{\cellcolor[HTML]{ECF4FF}s$_{33}$} & \multicolumn{1}{c}{\cellcolor[HTML]{ECF4FF}s$_{44}$} & \multicolumn{1}{c}{\cellcolor[HTML]{ECF4FF}s$_{66}$} \\ \cline{2-7} 
\multirow{-2}{*}{\begin{tabular}[c]{@{}c@{}}Elastic Compliance Constants\\ (10$^{-12}$ m$^2$/N)\end{tabular}} 
& \multicolumn{1}{c}{9.75} & -2.56 & \multicolumn{1}{c}{-3.79} & 10.36 & \multicolumn{1}{c}{15.88} & 15.92 \\ \hline 
& \multicolumn{2}{c}{\cellcolor[HTML]{ECF4FF}e$_{11}$} & \multicolumn{2}{c}{\cellcolor[HTML]{ECF4FF}e$_{22}$} & \multicolumn{2}{c}{\cellcolor[HTML]{ECF4FF}e$_{33}$} \\ \cline{2-7} 
\multirow{-2}{*}{Relative Permittivity} & \multicolumn{2}{c}{12550} & \multicolumn{2}{c}{12550} & \multicolumn{2}{c}{775} \\ \hline
\end{tabular}
\caption{Piezoelectric constants for PMN-PT used in the COMSOL simulations for strain, using the strain-charge form, from the literature \cite{PMN-PTconts}. }
\label{table:1}
\end{table}

\begin{figure}[h]
    \includegraphics[width=0.29\textwidth]{figures/Supplemental_Fig/New SI/raw_pos15V_XX_draft2.png}
    \includegraphics[width=0.29\textwidth]{figures/Supplemental_Fig/New SI/raw_pos15V_XY_draft2.png}
    \includegraphics[width=0.29\textwidth]{figures/Supplemental_Fig/New SI/raw_pos15V_YY_draft2.png}
    \includegraphics[width=0.08\textwidth]{figures/Supplemental_Fig/New SI/rawstrainaxis.png}
  \hfill
    \caption{Cross-sectional images of the components of the strain tensor in the $xyz$ frame, calculated with $+15$\,V applied to the side gates.}
    \label{fig:rawcompstraintensor}
\end{figure}

For the applied gate voltage of +15V, the strain was calculated in 3D space of the device, as shown in Figure~\ref{fig:rawcompstraintensor}.
To get the calculated strain of the surface onto the Fe$_{0.75}$Co$_{0.25}$ layer, the area under the channel was averaged and the strain tensor for $+15V$ was found to be

\begin{center}
$\varepsilon_{ij} =
\begin{bmatrix}
-1.9\times10^{-4} & -1.7\times10^{-4} & -0.04\times10^{-4} \\ -1.7\times10^{-4} & -1.9\times10^{-4} & 0.2\times10^{-4} \\ -0.04\times10^{-4} & 0.2\times10^{-4} & 16\times10^{-4}
\end{bmatrix}$
\end{center}
for the $+15$\,V applied. The shear strain along XZ and YZ are mainly on the edges of the current channel, and when averaging including the edges they are very small, and are more likely to be due to boundary conditions of the calculation. Thus, they are disregarded. 



To better understand the strain applied to the device, a rotational transformation was applied to the simulated strain to find the strain in the $x'y'z'$ frame. 
The rotated strain shows that main component of in-plane strain is $\epsilon'_{22}$, in the $y'$ direction between the side gates of the device. 

\begin{figure}[h]
    \includegraphics[width=0.29\textwidth]{figures/Supplemental_Fig/New SI/rotated_pos15V_XX_draft3.png}
    \includegraphics[width=0.29\textwidth]{figures/Supplemental_Fig/New SI/rotated_pos15V_XY_draft3.png}
    \includegraphics[width=0.29\textwidth]{figures/Supplemental_Fig/New SI/rotated_pos15V_YY_draft3.png}
    \includegraphics[width=0.08\textwidth]{figures/Supplemental_Fig/New SI/strainaxis.png}
  \hfill
    \caption{Cross-sectional images of the components of the strain tensor in the $x'y'z'$ frame, with $+15$\,V applied to the side gates.}
    \label{fig:rotatedcompstraintensor}
\end{figure}

The calculated strain underneath the device channel becomes:

\begin{center}
$\varepsilon'_{ij} =
\begin{bmatrix}
-0.19\times10^{-4} & 0 & 0 \\ 0 & -3.69\times10^{-4} & 0 \\ 0 & 0 & 16\times10^{-4}
\end{bmatrix}$
\end{center}




\subsection{Magnetoelastic Anisotropy}

We investigate whether the calculated strain is consistent with the observed change in the in-plane uniaxial anisotropy $K_u$ as a function of gate voltage. The magnetoelastic anisotropy in cubic materials such as Fe$_{0.75}$Co$_{0.25}$ is given by \cite{becker_ferromagnetismus_1939,wedler_magnetoelastic_2000,Sander2021}

\begin{align*}
E_{me} = \,\,&B_{1} \left[ \varepsilon'_{11}\left({m'_{1}}^{2}-\scriptstyle\frac{1}{3}\right) + \varepsilon'_{22}\left( {m'_{2}}^{2}-\scriptstyle\frac{1}{3}\right)
+  \varepsilon'_{33}\left({m'_{3}}^{2}-\scriptstyle\frac{1}{3}\right) \right] \\
&+ B_{2} \left[  \varepsilon'_{12}\left(m'_{1} m'_{2}\right) + \varepsilon'_{23}\left(m'_{2} m'_{3}\right) + \varepsilon'_{31}\left(m'_{3} m'_{1}\right) \right]
\end{align*}
where $B_1$ and $B_2$ are the magnetoelastic constants and 
$\displaystyle \vec{m}=\frac{\vec{M}}{|\vec{M}|}$ is the unit vector of the magnetization. The primes indicate that the vector and tensor components are for the coordinates of the Fe$_{0.75}$Co$_{0.25}$ cubic unit cell, i.e.~$(x', y', z')$, and the numerical indices correspond to $1\leftrightarrow x'$, $2\leftrightarrow y'$, and $3\leftrightarrow z'$.
As shown in Figure \ref{fig:piezosim_drawingdim}, $x'$ lies along the device channel.
Because the COMSOL simulation yields $\varepsilon'_{12} = 0$ and the magnetization is in-plane ($m'_3=0$), this reduces to
\begin{equation}
    E_{me} = B_{1} \left[ \varepsilon'_{11}\left({m'_{1}}^{2}-\scriptstyle\frac{1}{3}\right) + \varepsilon'_{22}\left( {m'_{2}}^{2}-\scriptstyle\frac{1}{3}\right)  \right]
    \label{eq:EvsM}
\end{equation}

In terms of the angle $\theta_M$ measured from the channel axis ($x'$), the magnetoelastic energy becomes
\begin{equation*}
 E_{me} = B_{1} \left( \varepsilon'_{22}-\varepsilon'_{11}\right) \sin^{2}{\theta_M} + \textrm{constant}
\end{equation*}
Since the in-plane uniaxial anisotropy coefficient $K_u$ is defined by $E=K_u \sin^2{\theta_M}$, the magnetoelastic contribution to the in-plane uniaxial anisotropy is given by
\begin{equation}
    K_{u}^{me} = B_1 \left( \varepsilon'_{22}-\varepsilon'_{11}\right)
    \label{eq:B1compact}
\end{equation}

The value of $B_1$ for Fe is $-3.44$ MJ/m$^{3}$ and is negative for FeCo alloys (Ref.~\cite{serizawa20191905R003} shows positive magnetostriction constants $\lambda_{100}$ for FeCo alloys, which corresponds to negative $B_1$ since $B_1=-\frac{3}{2}\lambda_{100}(c_{11}-c_{12})$
 \cite{Sander2021} and $c_{11}-c_{12}$ is positive \cite{hossain_systematic_2020}). 
Intuitively, this means with uniaxial tensile strain, the easy axis of the magnetoelastic energy will lie along the direction of elongation. 
This is most easily seen from equation~\ref{eq:EvsM}. For elongation along the $x'$ axis ($\varepsilon'_{11}>0)$, the overall negative coefficient ($B_1 \varepsilon'_{11}$) for ${m'_1}^2$ causes the energy to be lowered when $\vec{m}$ rotates to orient along the $x'$ axis.

 The overall in-plane uniaxial anisotropy $K_u$ consists of a constant term arising from magnetic shape anisotropy ($K_U^{msa}$) and the magnetoelastic contribution. Since the shape anisotropy favors alignment of the magnetization along the channel axis, this corresponds to a positive value of $K_U^{msa}$. The overall anisotropy is thus given by 
\begin{equation}
    K_u = K_u^{msa} +K_u^{me} = K_u^{msa}+B_1\left(\varepsilon'_{22}-\varepsilon'_{11}\right)
    \label{eq:total_Ku}
\end{equation}


This relationship allows us to estimate $B_1$ using values of $\varepsilon'_{22}-\varepsilon'_{11}$ from the COMSOL simulation and the variation of $K_{U}$ vs.~gate voltage from the experimental data in Figure 6a from the main text. The shape of the $K_u$ vs.~gate voltage in Figure 6a follows a quasi-linear shape on both the up sweep and down sweep between $+15$\,V and $-15$\,V. This is in contrast to hysteretic butterfly-shaped curves of strain vs.~applied electric field observed in bottom-gated ferromagnet/PMN-PT(001) samples \cite{schoffmann_strain_2022}. The main difference is that the butterfly shape occurs when a full ferroelectric hysteresis loop is traversed, while a quasi-linear shape occurs when a minor loop is traversed (i.e.~the applied electric field remains below the coercive field). From the data in Figure 6a, we quantify the variation of $K_u$ with gate voltage by its average slope between $-15$\,V and $+15$\,V: $\frac{\Delta K_u}{\Delta V_g} = \left( \frac{3.1 \times 10^4\,\text{J/m}^3 - 5.5 \times 10^4\,\text{ J/m}^3}{30\,\text{V}} \right) = -800\,\frac{\text{J}}{\text{m}^3 \text{V}}$. Because the sign of the slope depends on the remnant polarization state which is unknown, we can only consider the magnitude of the slope, $\left| \frac{\Delta K_u}{\Delta V_g}  \right| = 800\,\frac{\text{J}}{\text{m}^3 \text{V}}$.


Using equation \ref{eq:B1compact} and the strain calculated in COMSOL, we determine the magnitude of the magnetoelastic constant to be $\left| B_1 \right| = \left| \frac{K_u^{me}}{\varepsilon'_{22}-\varepsilon'_{11}} \right| = \left| \frac{(\Delta K_u/\Delta V_g)*(V_g)}{\varepsilon'_{22}-\varepsilon'_{11}} \right| = \left| \frac{(800\frac{\text{J}}{\text{m}^3 \text{V}})*(15\,\text{V})}{-3.69\times10^{-4}+0.19\times10^{-4} \frac{J}{m^3}} \right| = 35$\,MJ/m$^3$. 
Because the sign of $B_1$ is negative for FeCo \cite{serizawa20191905R003}, our anisotropy data and strain simulations yield a value of $B_1 = -35$\,MJ/m$^3$.

We now consider whether the extracted value of $-35$ MJ/m$^3$ is reasonable for the $B_1$ of Fe$_{0.75}$Co$_{0.25}$. A recent measurement of magnetostriction of single-crystalline Fe$_{0.70}$Co$_{0.30}$ films on MgO(100) reports a magnetostriction constant $\lambda_{100}$ of $+234\times10^{-6}$ \cite{serizawa20191905R003}. This can be converted to $B_1$ by the formula \cite{Sander2021}
\begin{equation}
    B_1=-\frac{3}{2}\lambda_{100}(c_{11}-c_{12})
    \label{eq:B1fromlambda}
\end{equation}
where $c_{11}$ and $c_{12}$ are elastic constants.  
In lieu of a direct measurement for Fe$_{0.70}$Co$_{0.30}$, we estimate $c_{11}-c_{12}$ as follows. 
Theoretical calculations of FeCo alloys show that $c_{11}-c_{12}$ varies slightly with composition and has values of 137 GPa for Fe and 145 GPa for Fe$_{0.75}$Co$_{0.25}$ \cite{hossain_systematic_2020}. Meanwhile, experimental measurements on monocrystalline Fe samples yield a value of 96 GPa \cite{adams_elastic_2006}. Therefore, we conservatively estimate that $c_{11}-c_{12}$ is between 96 and 150 GPa. Using equation \ref{eq:B1fromlambda}, this produces a value of $B_1$ between $-34$ and $-53$ MJ/m$^3$. This shows that the value of $B_1=-35$ MJ/m$^3$ estimated from our experimental data and COMSOL simulations is quite reasonable. Thus, this analysis provides additional support that the magnetoelectric coupling originates from a strain-mediated effect.


\bibliography{refs.bib}